\documentclass[12pt,preprint]{aastex631}
\usepackage{natbib}
\usepackage{xcolor}

\newcommand{\eps}[1]{\mbox{log~$\epsilon$(#1)}} 
\newcommand\species[2]{#1~{\sc #2}}
\newcommand\iso[2]{$^{\rm #1}$#2}

\def\eg{\mbox{e.g.}}

\def\teff{\mbox{T$_{\rm eff}$}}
\def\logg{\mbox{log~{\it g}}}
\def\vmicro{\mbox{$\xi_{\rm t}$}}
\def\kmsec{\mbox{km~s$^{\rm -1}$}}

\shorttitle{Metal-poor Stellar Fe-Group Abundances}
\shortauthors{Sneden et al.}

\begin{document}

\title{Iron-Peak Element Abundances in Warm Very Metal-Poor Stars}

\author[0000-0002-3456-5929]{Christopher Sneden}
\affiliation{Department of Astronomy and McDonald Observatory, The University 
             of Texas, Austin, TX 78712; chris@verdi.as.utexas.edu}
\author[0000-0002-8468-9532]{Ann Merchant  Boesgaard}
\affiliation{Institute for Astronomy, University of Hawai’i at Manoa, 
             2680 Woodlawn Drive, Honolulu, HI 96822, USA; boes@ifa.hawaii.edu}
\author[0000-0002-6779-3813]{John J. Cowan}
\affiliation{Homer L. Dodge Department of Physics and Astronomy, University 
             of Oklahoma, Norman, OK 73019; jjcowan1@ou.edu}
\author[0000-0001-5107-8930]{Ian U. Roederer}
\affiliation{Department of Astronomy, University of Michigan, 
             1085 S. University Ave., Ann Arbor, MI 48109, USA; iur@umich.edu}
\affiliation{Joint Institute for Nuclear Astrophysics -- Center for the
Evolution of the Elements (JINA-CEE), USA}
\author[0000-0001-8582-0910]{Elizabeth A. Den Hartog}
\affiliation{Department of Physics, University of Wisconsin-Madison,
             1150 University Ave., Madison, WI 53706: eadenhar@wisc.edu}
\author[0000-0001-5579-9233]{James E. Lawler}
\affiliation{Department of Physics, University of Wisconsin-Madison,
             1150 University Ave., Madison, WI 53706}
\affiliation{Deceased January 29, 2023}

\begin{abstract}
We have derived new detailed abundances of Mg, Ca, and the Fe-group elements 
Sc through Zn (Z~=~21$-$30) for 37 main sequence turnoff very metal-poor stars 
([Fe/H]~$\lesssim$~$-$2.1).
We analyzed Keck HIRES optical and near-UV high signal-to-noise spectra 
originally gathered for a beryllium abundance survey.
Using typically $\sim$400 Fe-group lines with accurate laboratory transition
probabilities for each star, we have determined accurate LTE metallicities 
and abundance ratios for neutral and ionized species of the 10 Fe-group 
elements as well as $\alpha$ elements Mg and Ca.
We find good neutral/ion abundance agreement for the 6 elements that have
detectable transitions of both species in our stars in the 3100$-$5800~\AA\
range.
Earlier reports 
of correlated Sc-Ti-V relative overabundances are confirmed,
and appear to slowly increase with decreasing metallicity.
To this element trio we add Zn; it also appears to be increasingly 
overabundant in the lowest metallicity regimes.
Co appears to mimic the behavior of Zn, but issues surrounding its 
abundance reliability cloud its interpretation.
\end{abstract}

\section{INTRODUCTION\label{intro}}

Almost all elements in the Universe are formed in stars.
Elemental abundances in very low metallicity stars, among the oldest 
in the Galaxy and the 
Universe, provide important clues to early Galactic nucleosynthesis.  
Such observed abundances provide insight into the identities of the very 
first stars (long-since gone) and the nature of the nuclear processes that 
occurred in those stars (\eg, \citealt{curtis19}, \citealt{ebinger20}, 
\citealt{cowan21}).
These processes can occur in environments typical of supernovae (SNe) in 
core-collapse explosions from massive stars, or Type Ia explosions resulting 
from binary interactions.
Comparisons between the low metallicity stars and more metal-rich groups 
indicate changes in chemical evolution over time pointing to variations in 
stellar mass ranges and synthesis mechanisms over Galactic timescales. 
Such comparisons, however, require accurate abundance determinations, 
which in turn depend sensitively upon precise experimental atomic physics data 
and high-resolution, high signal-to-noise ratio (SNR) astronomical spectra.

Our earlier work in this area focused on the neutron-capture, particularly 
Rare Earth, elemental abundances in stars. 
Those efforts involved obtaining large amounts of experimental atomic data 
(line transition probabilities and isotopic/hyperfine sub-component
parameters) from the Wisconsin physics group.
These new experimental data, 
applied to ground-based and space-based stellar spectra,
led to precise elemental abundances of a large number of neutron-capture 
elements in metal-poor stars (\eg, \citealt{lawler09}, \citealt{sneden09},
\citealt{holmbeck18}, \citealt{roederer22} and references therein).

More recently we have concentrated on iron-group elements, Z~=~21$-$30 
(see \citealt{sneden16}, \citealt{lawler17}, \citealt{wood18}, 
\citealt{lawler18}, \citealt{lawler19}, \citealt{denhartog19},  
\citealt{cowan20}).
Many of these studies used the bright well-known metal-poor
([Fe/H]~$\sim$~$-$2.2)\footnote{
We adopt the standard spectroscopic notation \citep{wallerstein59} that for 
elements A and B,
[A/B] $\equiv$ log$_{\rm 10}$(N$_{\rm A}$/N$_{\rm B}$)$_{\star}$ $-$
log$_{\rm 10}$(N$_{\rm A}$/N$_{\rm B}$)$_{\odot}$.
Often in this paper we will expand the [A/B] notation to the species level,
\eg, [\species{A}{i}/\species{B}{i}] or [\species{A}{ii}/\species{B}{ii}], 
signifying elemental abundances determined from neutral or ionized species.
We equate metallicity with the stellar [Fe/H] value, and compute it as
the mean of [\species{Fe}{i}/H] and [\species{Fe}{ii}/H] abundances.
To form the differential [A/B] quantities we use the solar abundances
of \cite{asplund09}.
Finally, we use the definition 
\eps{A} $\equiv$ log$_{\rm 10}$(N$_{\rm A}$/N$_{\rm H}$) + 12.0;
 \eps{\species{X}{i}} or \eps{\species{X}{ii}} are to be understood as  
an elemental abundance determined from the named species.}
main sequence turnoff star HD 84937, which has well-known atmospheric 
parameters and large amounts of high-resolution spectroscopic data.
In \citeauthor{cowan20} Fe-group abundances were determined in 3
main-sequence turnoff stars ([Fe/H]~$\sim$~$-$3) with 
HST/STIS vacuum-$UV$ (2300$-$3050~\AA) high resolution spectra.
In that study we were able to derive abundances from both neutral and ionized
species for 6 out of the 10 Fe-group elements.

Unfortunately, only a few very metal-poor stars have vacuum $UV$ high 
resolution spectra, and that number is unlikely to grow much in the future.
From basic atomic structure considerations, for most metallic species the
number of transitions and their strengths rise rapidly with decreasing
wavelength (\eg, Figure~2 in \citealt{lawler17} and other examples in our
laboratory transition paper series).
Therefore, many of the Fe-group species best studied in the vacuum-$UV$
also have large numbers of transitions available in the near-$UV$, 
$\lambda\lambda$3000$-$4000~\AA.
\cite{boesgaard11} (hereafter called B11) derived Be and alpha element
abundances for 117 main-sequence stars in the disk and halo 
($-$3.5~$\leq$~[Fe/H]~$\leq$~$-$0.5).
For Be they studied the \species{Be}{ii} resonance doublet at 
3130.4, 3131.1~\AA.
Gathering high SNR spectra at these wavelengths ensured that we could study
Fe-group transitions nearly to the atmospheric ozone cutoff.

Here we present new abundances of Fe-group elements in 37 very metal-poor 
main sequence turnoff stars ([Fe/H]~$\leq$ $-$2.0), from our analyses
of B11 Keck HIRES spectra.
In \S\ref{specdata} we introduce the high-resolution spectroscopic
data set, followed by the abundance analysis in \S\ref{abanalysis}.
Interpretation of the abundance trends is given in  \S\ref{discussion}.
Finally, a summary and conclusions are detailed in \S\ref{sum}.

\section{SPECTROSCOPIC DATA\label{specdata}}

The B11 Be abundance survey required high SNR data near 3100~\AA.
This spectral region suffers significant telluric
extinction, so only relatively $UV$-bright metal-poor stars were available
for the \citeauthor{boesgaard11} study.
This restriction generally ruled out metal-poor red giants, which have 
steeply declining fluxes in the near-$UV$.
Additionally, the deep atmospheres of red giants produce very strong-lined
spectra even for metallicities [Fe/H]~$<$~$-$2, rendering abundance analyses
in the near-$UV$ very difficult.
Therefore the B11 sample was limited to warm main 
sequence turnoff and subgiant stars: 
5500~K~$\lesssim$ \teff~$\lesssim$ 6500~K, and
3.0~ $\lesssim$ \logg~$\lesssim$ 4.8.

The spectra were all obtained with Keck~I/HIRES \citep{vogt94}, which
yielded spectral resolving power 
$R$~$\equiv$ $\lambda/\Delta\lambda$ $\simeq$ 42,000 in the near-$UV$.  
These echelle spectra have near-continuous wavelength coverage in the interval
3050~\AA~$\lesssim$~$\lambda$~$\lesssim$ 5900~\AA.
For the complete data set B11 estimated the signal-to-noise to be  
$\langle$SNR$\rangle$ = 106 at 3130~\AA.
The $SNR$ increases rapidly toward longer wavelengths, typically exceeding
values of 200 near 4000~\AA\ and 300 near 5000~\AA.
The $SNR$ ratio is not an issue in our abundance analyses.

\begin{figure}
\epsscale{0.7}
\plotone{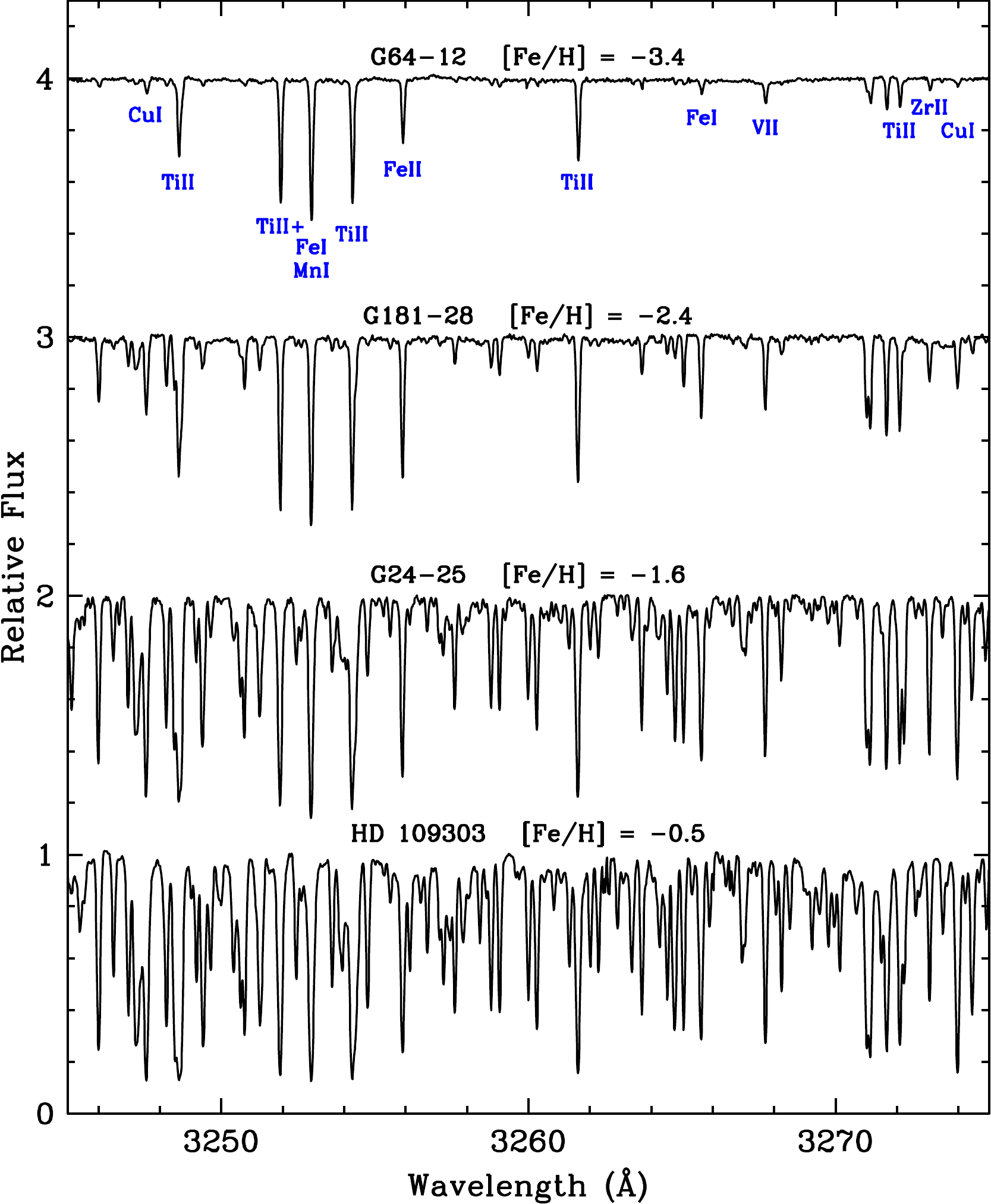}
\caption{
\label{fig1} \footnotesize
   Sample Keck~I/HIRES spectra of four stars from the B11 survey.
   This small region illustrates the near-UV line density in main sequence 
   turnoff stars of different metallicities.
   Stars G64-12 and G181-28 are part of the present study, while stars 
   G24-25 and HD~109303 are too ``metal-rich'' to be included here.
   The metallicities quoted in the figure legend are from B11 Table~2.
   The relative flux scale is correct for HD 109303, and the other spectra 
   have been shifted successively by additive constants of 1.0 for display
   purposes.
   Major absorption features have been identified for the G64-12 spectrum.
}
\end{figure}

A more significant concern is the natural complexity of the program star
near-$UV$ spectra at high metallicities.
In Figure~\ref{fig1} we show a small near-$UV$ wavelength region of
interest to this investigation.
Note especially in this figure the two \species{Cu}{i} resonance 
lines, detectable even in the lowest metallicity stars.
Four stars with metallicities increasing by about a factor of 10 each step
from top to bottom in the figure are displayed.
Inspection of these spectra suggests that while continuum points can be easily
established for stars with [Fe/H]~$\lesssim$~$-$2.0, this determination becomes
challenging by [Fe/H]~$\gtrsim$~$-$1.5 and practically impossible for
[Fe/H]~$\gtrsim$~$-$1.0.

Our astrophysical goal was to investigate the relative abundances of Fe-group
elements in very metal-poor stars, using very large numbers of neutral and
ionized lines with recent laboratory transition data.  
To accomplish this task for many stars, examining each transition for 
its abundance utility in each star, we used equivalent width ($EW$) 
measurements instead of synthetic/observed spectrum matches.
The near-UV absorption line crowding displayed in the higher metallicity 
stars of Figure~\ref{fig1} left relatively few unblended spectral 
features in them for analysis.
Therefore we limited our program star list to the B11 stars with
[Fe/H]~$\lesssim$~$-$2.1, adding one star with [Fe/H]~$\simeq$~$-$1.4
to test our analytical methods in a higher metallicity regime.

Final reduction steps were accomplished with IRAF\footnote{
https://iraf-community.github.io/}
software \citep{tody86,tody93}, including clipping of radiation events with 
task $lineclean$ and concatenation of spectral orders with task $scombine$.
Then the spectra were divided into 100~\AA\ segments.
These spectrum pieces were subjected to detailed continuum normalization
with spline functions in the specialized software package $SPECTRE$
\citep{fitzpatrick87}\footnote{
Available at http://www.as.utexas.edu/$\sim$chris/spectre.html}, as well as
smoothing with 2-pixel Gaussian functions.

\section{ABUNDANCE ANALYSIS\label{abanalysis}}

All abundance computations were performed with the current version of the
plane-parallel LTE line analysis code MOOG \citep{sneden73}\footnote{
Available at http://www.as.utexas.edu/$\sim$chris/moog.html}.
The program stars are all warm, high gravity stars, with continuum 
opacities dominated by H$^-$ even at the near-$UV$ wavelengths of this
study.
Therefore no special computations \citep{sobeck11} were required to account
for continuum Rayleigh scattering, which we confirmed with abundance
tests.

We employed more than about 350 Fe-group lines for each program star, 
which made synthetic spectrum analyses impractical, as discussed in 
\S\ref{specdata}.  
We relied almost exclusively on $EW$ measurements,
which limited the line lists to relatively unblended transitions
with nearby well-defined continuum points.
We used $SPECTRE$ for semi-automated $EW$ measurements, with visual
inspection of each line profile.
This procedure allowed for elimination of very weak lines and those with
distorted profiles, and for re-measurement of very strong lines that display
obvious Voigt profile wings.
In the weak-line domain we usually discarded lines with reduced widths 
log~$RW$~$\equiv$ log($EW$/$\lambda$)~$\lesssim$ $-$6.3 ($EW$~$\sim$~2~m\AA\
at 4000~\AA) and only retained lines with log~$RW$~$<$ $-$6.0 if their line
profiles were clean and not affected by continuum noise fluctuations.
For strong lines we discarded lines with log~$RW$~$\gtrsim$ -4.5 
($EW$~$\sim$125~m\AA\ at 4000~\AA).
Lines stronger than this limit are on the damping part of the curve-of-growth.
Abundances derived from them depend strongly on microturbulent velocity
and damping parameters.
Our $EW$s are available in the Zenodo database.\footnote{
https://zenodo.org/record/7820255}

\subsection{Atomic Transitions\label{transitions}}

Discussions of line selection and laboratory data have appeared in 
\cite{sneden16}, and \cite{cowan20}, often with comments on the laboratory 
studies that will be cited here on individual Fe-group species.
Only lines from these papers are employed in our study.
Data for these transitions may be found in the $linemake$ on-line
facility \citep{placco21}.\footnote{
https://github.com/vmplacco/linemake}

Most lines were assumed to be single transitions, but those of odd-Z elements 
Sc, V, Mn, Co, and Cu have significant hyperfine substructure (hfs) that
can significantly broaden spectral features, leading to desaturation and
consequent decrease in derived abundances from strong lines.
The hfs substructures are especially large for \species{Co}{i}, with 
abundance corrections reaching $\sim$0.5~dex for the strongest transitions.
For all lines that have known hfs patterns and are strong enough to have any 
possibility of saturation ($RW$ $>$ $-$5.5), we derived abundances with
$MOOG$'s $EW$ option \textit{blends}.
Here we briefly summarize the lines used in our computations.
Issues arising in the abundance computations will be discussed in 
\S\ref{abunds}.

\textbf{Magnesium:}
There are no recent comprehensive transition probability studies of 
\species{Mg}{i} lines, but generally they are well-determined from past
atomic physics studies.
The NIST Atomic Spectra Database (NIST ASD, \citealt{kramida22}\footnote{
https://www.nist.gov/pml/atomic-spectra-database})
gives quality ratings of ``B'' ($<\pm$10\% uncertainty) for all lines of 
interest here, except for 4571.10~\AA (rating D, $<\pm$50\% uncertainty).
None of the Mg abundances in our study were determined only from the
$\lambda$4571 line.

\textbf{Calcium:}
This element was not considered in our earlier papers due to the lack of
recent laboratory work.
A new study by \cite{denhartog21} has derived transition probabilities
for \species{Ca}{i}, and we have adopted log($gf$) values from that work.
A few \species{Ca}{ii} lines that are weak enough for abundance analyses
appear in our spectra, notably those at $\lambda\lambda$3158.87, 3179.33, 
and 3706.03.
\citeauthor{denhartog21} showed that abundances from \species{Ca}{ii}
lines using transition probabilities from \cite{safronova11} were in
agreement with \species{Ca}{i} values for one metal-poor main sequence star,
HD~84937. 
We included these \species{Ca}{ii} transitions in the present study.

\textbf{Scandium:}
Laboratory transition probabilities and hfs patterns for \species{Sc}{ii} 
were taken from \cite{lawler19}.
No \species{Sc}{i} lines are strong enough for detection in metal-poor stars.
Hyperfine substructure does affect \species{Sc}{ii} and lowers the 
derived Sc abundances, but the effect is
modest for our stars, typically $\lesssim$0.03~dex

\textbf{Titanium:}
Neutral and ionized species of Ti have many available transitions in our
spectra.
Their lab data were taken from \cite{lawler13} and \cite{wood13}.

\textbf{Vanadium:}
Our abundances were mostly based on numerous transitions of \species{V}{ii}.
In a few stars a handful of extremely weak \species{V}{i} lines can be 
detected.
The transition data were adopted from \cite{lawler14} and \cite{wood14a}.
Full hfs calculations were necessary for the \species{V}{ii} transitions.

\textbf{Chromium:}
The laboratory transition data for \species{Cr}{i} and \species{Cr}{ii}
were taken from \cite{sobeck07} and \cite{lawler17}, respectively.    
Our abundances were determined from large line samples, on average 8 for
\species{Cr}{i} and 16 for \species{Cr}{ii}.
This permitted us to explore the known disagreement between Cr abundances
derived from neutral and ionized species (\eg, \citealt{kobayashi06},
\citealt{sobeck07}, \citealt{roederer14}).

\textbf{Manganese:}
Transition probabilities and hfs components for both
\species{Mn}{i} and \species{Mn}{ii} were taken from \cite{denhartog11}.    
Only about 5 lines each are available for neutral and ionized Mn species.
For \species{Mn}{i} often the derived abundances were based entirely on 
the resonance triplet at 4030.76, 4033.07, and 4034.49~\AA.
This set of lines has been responsible for most Mn abundance results in     
the literature of very metal-poor stars, and is known to yield much lower   
abundances than other neutral and ionized Mn lines.           
We will discuss this issue in \S\ref{abunds}.

\textbf{Iron:}
As in \cite{cowan20}, we limited our \species{Fe}{i} transition set to those
published in the collaborative laboratory effort of the Imperial College
London and the University of Wisconsin atomic physics groups:
\cite{ruffoni14}, \cite{denhartog14}, \cite{belmonte17}.
Recently \cite{denhartog19} determined new transition probabilities
for \species{Fe}{ii}.
Most of that laboratory effort concentrated on the very strong lines in the 
vacuum UV, and the $gf$ values at longer wavelengths combined new branching 
fractions and previous lifetime measurements.  
See \cite{denhartog19} \cite{cowan20} for further comments on this issue.

\textbf{Cobalt:}
Both \species{Co}{i} and \species{Co}{ii} have recent lab transition
probability and hfs analyses (\citealt{lawler15}, \citealt{lawler18}).
However, \species{Co}{ii} has strong transitions only in the vacuum~$UV$
($\lambda$~$<$~3050~\AA).
\cite{lawler18} report lab $gf$~values for six lines in the 3300$-$3600~\AA\
spectral region.  
Unfortunately they all are very weak in our program stars, and most of them 
are severely blended with transitions of other species.  
In the solar spectrum, the relatively unblended \species{Co}{ii} lines have 
$EW$~$\lesssim$ 30~m\AA\ \citep{moore66}.
In our metal-poor program stars their maximum $EW$ values should be no more 
than a few m\AA.
We verified their absence in each of our stars.

\textbf{Nickel:}
Ni abundances were derived exclusively from \species{Ni}{i} transitions,
with laboratory data taken from \cite{wood14b}.
All potentially detectable \species{Ni}{ii} lines occur in the vacuum~$UV$,
inaccessible for this study.
\cite{cowan20} had HST/STIS data for their small sample of turnoff stars
with [Fe/H]~$\sim$~$-$3, and the two Ni species yielded consistent abundances,
suggesting that the \species{Ni}{i} lines used here can be trusted to
yield reliable elemental abundances.
 
\begin{figure}
\epsscale{0.8}
\plotone{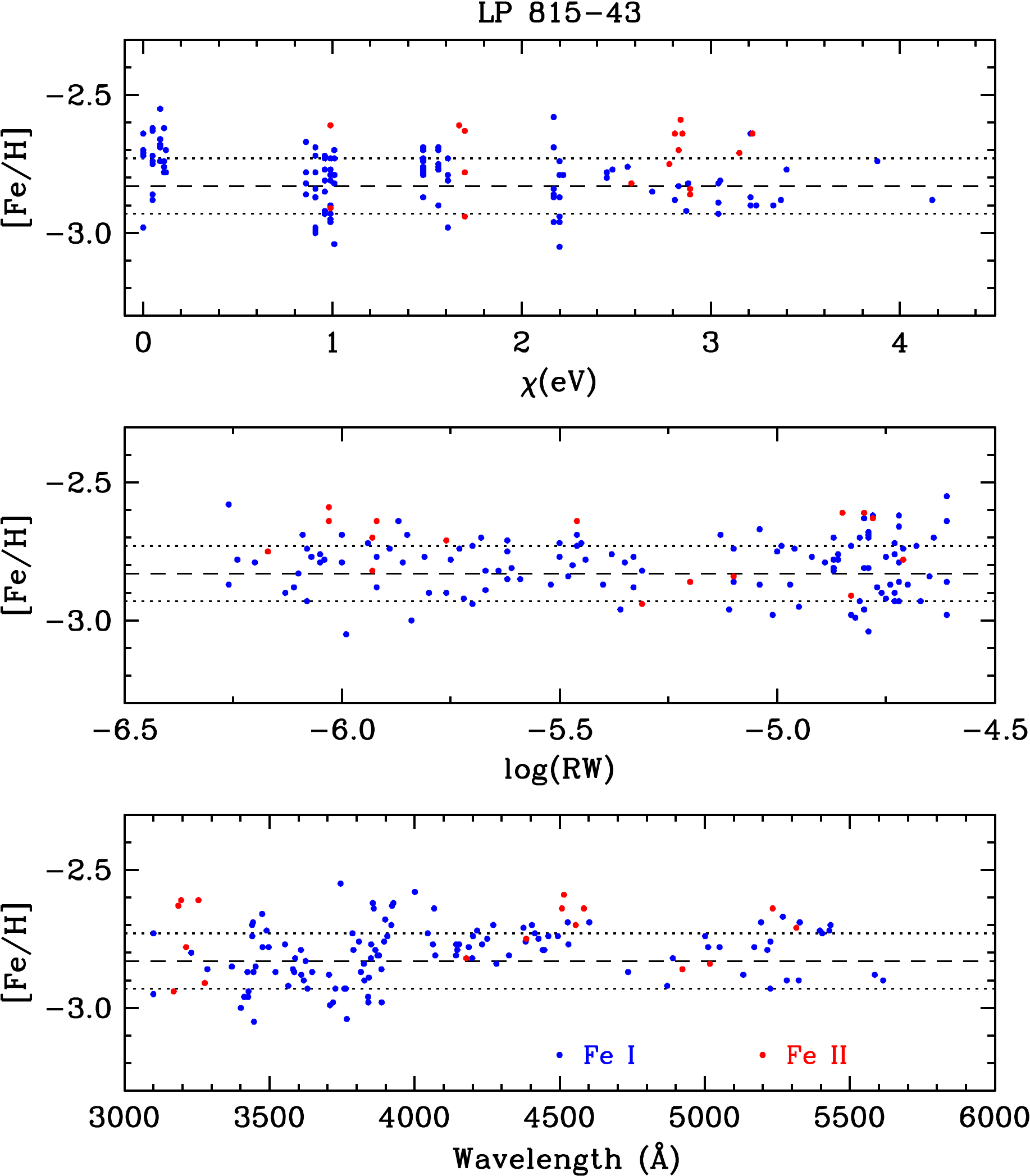}
\caption{
\label{fig2} \footnotesize                               
   Abundances from individual \species{Fe}{i} and \species{Fe}{ii} lines
   (blue and red points, respectively) in the program star LP~815-43.
   The three panels show these abundances as functions of excitation
   energy $\chi$, reduced width $RW$, and wavelength.
   In each panel the dashed line represents the mean abundance from 
   \species{Fe}{i}, and the dotted lines represent the sample standard
   deviations $\sigma$.
   The Fe abundances are those derived with a model atmosphere with
   \teff~=~6350~K, \logg~=~4.00, \vmicro~=~1.30~\kmsec, and model 
   metallicity [Fe/H]~=~$-$2.9.
}                                                            
\end{figure} 

\textbf{Copper:}
The atomic structure of this element gives rise only to a few \species{Cu}{i}
lines; \species{Cu}{ii} has no detectable lines in the optical spectral region.
Usually the \species{Cu}{i} resonance lines at 3247.5 and 3273.9~\AA\
were the only detectable transitions, with the 5105.5~\AA\ line available on
rare occasions.
There are no recent lab studies of this species, but the transition 
probabilities of the resonance lines are of high accuracy: the NIST database
rates them as ``AA'' ($\leq$1\% uncertainty).
Abundance computations including hfs were performed for these lines
using the $gf$ values recommended in the NIST ASD \citep{kramida19}\footnote{
Atomic Spectra Database of the National Institute of Standards and Technology;
https://physics.nist.gov/PhysRefData/ASD/lines\_form.html} website.

\textbf{Zinc:}
We could only detect the \species{Zn}{i} near-UV high-excitation lines at 
3302.6, 3345.0~\AA\ to supplement the optical lines at 4722.16 and 4810.54~\AA\
that have been employed in nearly all previous studies in metal-poor stars.
All of these lines are extremely weak, and abundances were derived for them
from both $EW$ and synthetic spectrum analyses.
The near-UV lines have multiple components.
Each of them has one strong member; the weaker components are split redward by 
$\gtrsim$0.4~\AA\ and are blended and barely detectable in our spectra.

\subsection{Model Atmospheres\label{modelatm}}

Initial model atmospheric parameters \teff, \logg, [Fe/H] metallicity,
and \vmicro\ were taken from Table~2 of B11.
These values were used to generate interpolated models from the 
\cite{kurucz11,kurucz18} alpha-enhanced model grid\footnote{
http://kurucz.harvard.edu/grids.html} using software kindly provided by
Andrew McWilliam and Inese Ivans.

We determined abundances with these models and assessed the assumed model
parameters with standard criteria:
for \teff, that there be no significant abundance differences between
low and high excitation lines of \species{Fe}{i}; for \vmicro, that there
be no differences between weak and strong lines for \species{Fe}{i} and
\species{Ni}{i}; for \logg, that the mean neutral and ionized abundances
of Ca, Ti, and Fe agree; and for metallicity, that the model [Fe/H] was
similar to the derived [Fe/H]. 
In comparing abundances from weak and strong lines, from low and high
excitation lines, and from neutral and ionized lines the mean abundance 
differences were held to less than 0.10~dex whenever possible.
In Figure~\ref{fig2} we show a typical example of this process applied
to the neutral and ionized Fe lines.  

In Table~\ref{tab-models} we list the derived model parameters and 
metallicities determined from both \species{Fe}{i} and \species{Fe}{ii}
transitions.
In general our new parameters are close to those derived by B11.
Defining $\Delta$A~$\equiv$ A$_{this~work}$ $-$ A$_{B11}$, 
for any parameter A, we found
$\langle\Delta$\teff$\rangle$~= 80~$\pm$~23~K ($\sigma$~=~137~K);
$\langle\Delta$\logg$\rangle$~= $-$0.05~$\pm$~0.05 ($\sigma$~=~0.33);
$\langle\Delta$\vmicro$\rangle$~= $-$0.15~$\pm$~0.02~\kmsec\
        ($\sigma$~=~0.14~\kmsec); and
$\langle\Delta$[Fe/H]$\rangle$~= $-$0.10~$\pm$~0.02 ($\sigma$~=~0.13).

Only one star merits comment in this atmospheric parameter comparison,
BD~$-$10 388.
Our analysis of this star yielded somewhat different values compared to B11.
Our parameters (\teff,\logg, [Fe/H]) = (6350,4.00,$-$2.43) put BD~$-$10 388
essentially at the main sequence turnoff, while the B11 parameters
(5768~K,3.04,$-$2.79) suggest subgiant status.
Other studies yield mixed parameters, \eg,
(6009,4.14,$-$2.33, \citealt{reddy08}), or
(5931,3.77,$-$2.62, \citealt{bensby18}).
The Gaia photometric-based parameters \citep{GAIA16,GAIA22} 
(6275,4.05,$-$2.22)) are in good agreement with our values.
BD~$-$10 388 deserves further investigation in the future, but excluding
it from our comparison with B11 yields little change in the mean
results and smaller scatter:
$\langle\Delta$\teff$\rangle$~= 66~$\pm$~18~K ($\sigma$~=~109~K);
$\langle\Delta$\logg$\rangle$~= $-$0.07~$\pm$~0.05 ($\sigma$~=~0.28);
$\langle\Delta$(\vmicro$\rangle$~= $-$0.16~$\pm$~0.02~\kmsec\
         ($\sigma$~=~0.14~\kmsec); and
$\langle\Delta$[Fe/H]$\rangle$~= $-$0.11~$\pm$~0.02 ($\sigma$~=~0.10).

\subsection{Abundances\label{abunds}}

\begin{figure}
\epsscale{0.8}
\plotone{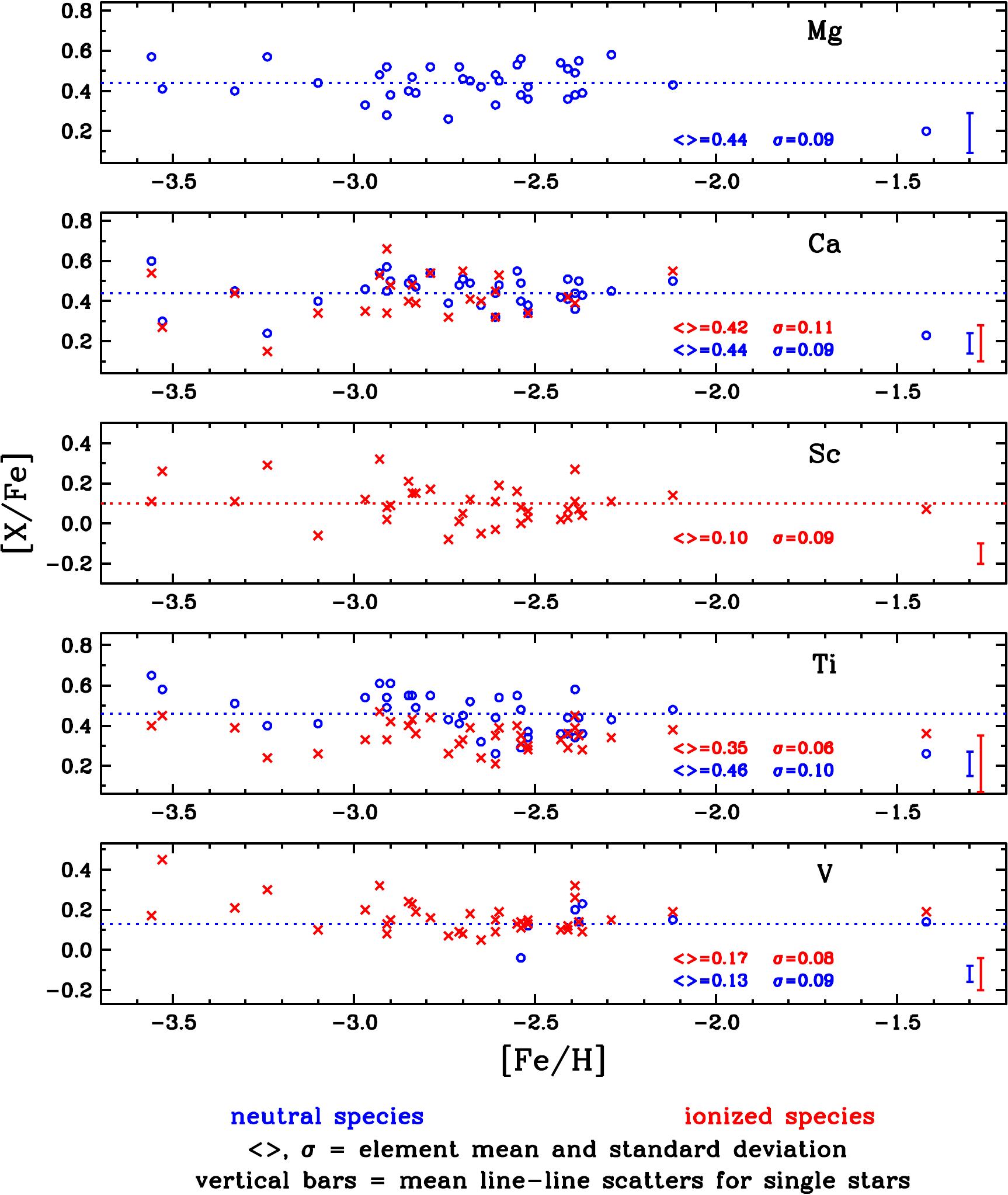}
\caption{
\label{fig3} \footnotesize
   Elemental abundance ratios [X/Fe] as functions of [Fe/H] metallicity for 
   light elements Mg, Ca, and the 3 lightest Fe-group elements Sc, Ti, and V.
   Blue colors, lines and open circles are for neutral-species transitions,
   while red and X-symbols are for ionized species transitions.
   Vertical bars in the lower corners indicate the mean line-to-line abundance
   scatters for individual abundances.
   The quoted statistics refer to the mean [X/Fe] value for a species for
   all stars, and $\sigma$ is its sample deviation.
}
\end{figure}

\begin{figure}
\epsscale{0.8}
\plotone{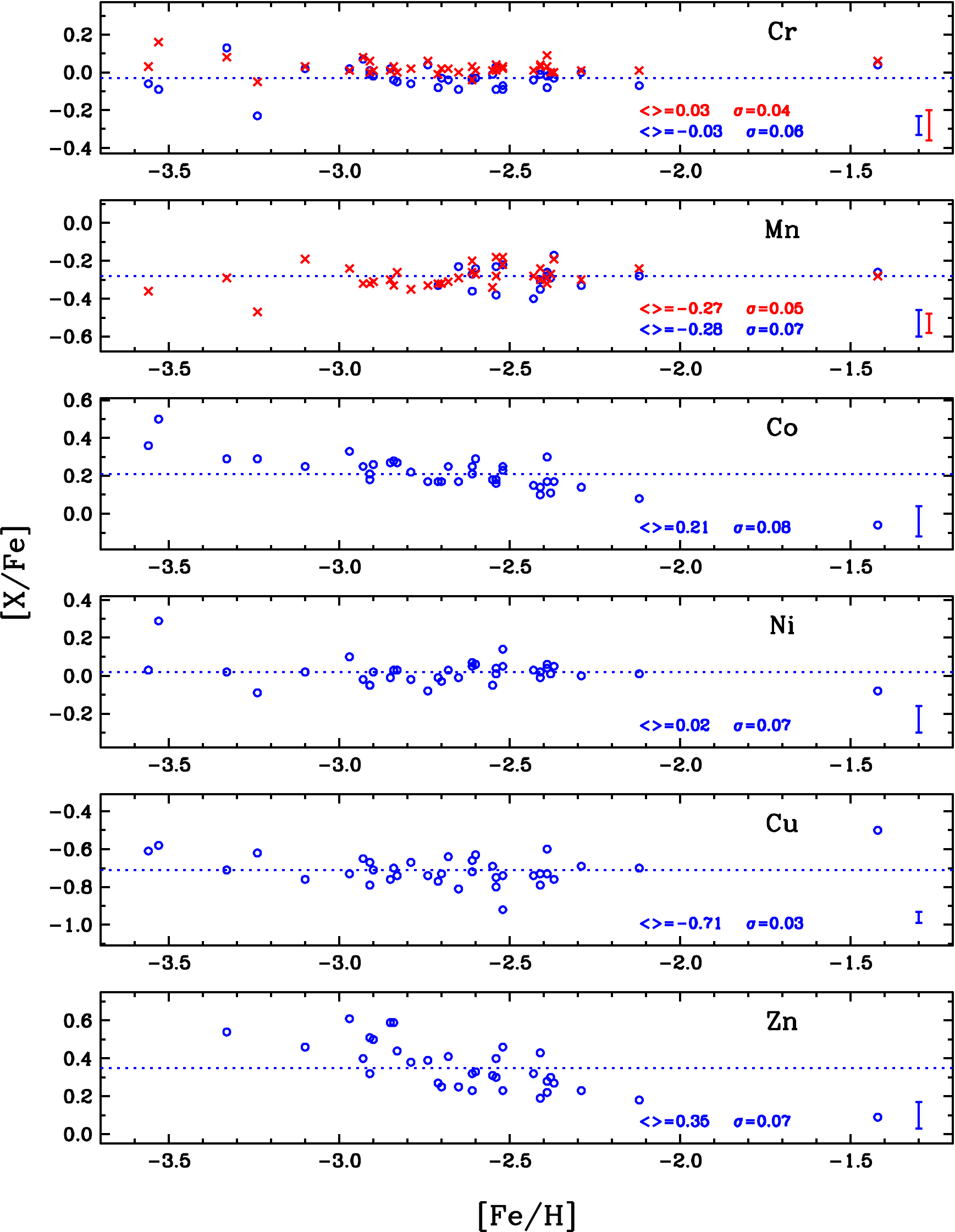}
\caption{
\label{fig4} \footnotesize
   Elemental abundance ratios [X/Fe] as functions of [Fe/H] metallicity for 
   heavier Fe-group elements Cr, Mn, Co, Ni, Cu, and Zn.
   Colors, symbols, and lines are as in Figure~\ref{fig3}.
}
\end{figure}

\begin{figure}
\epsscale{0.7}
\plotone{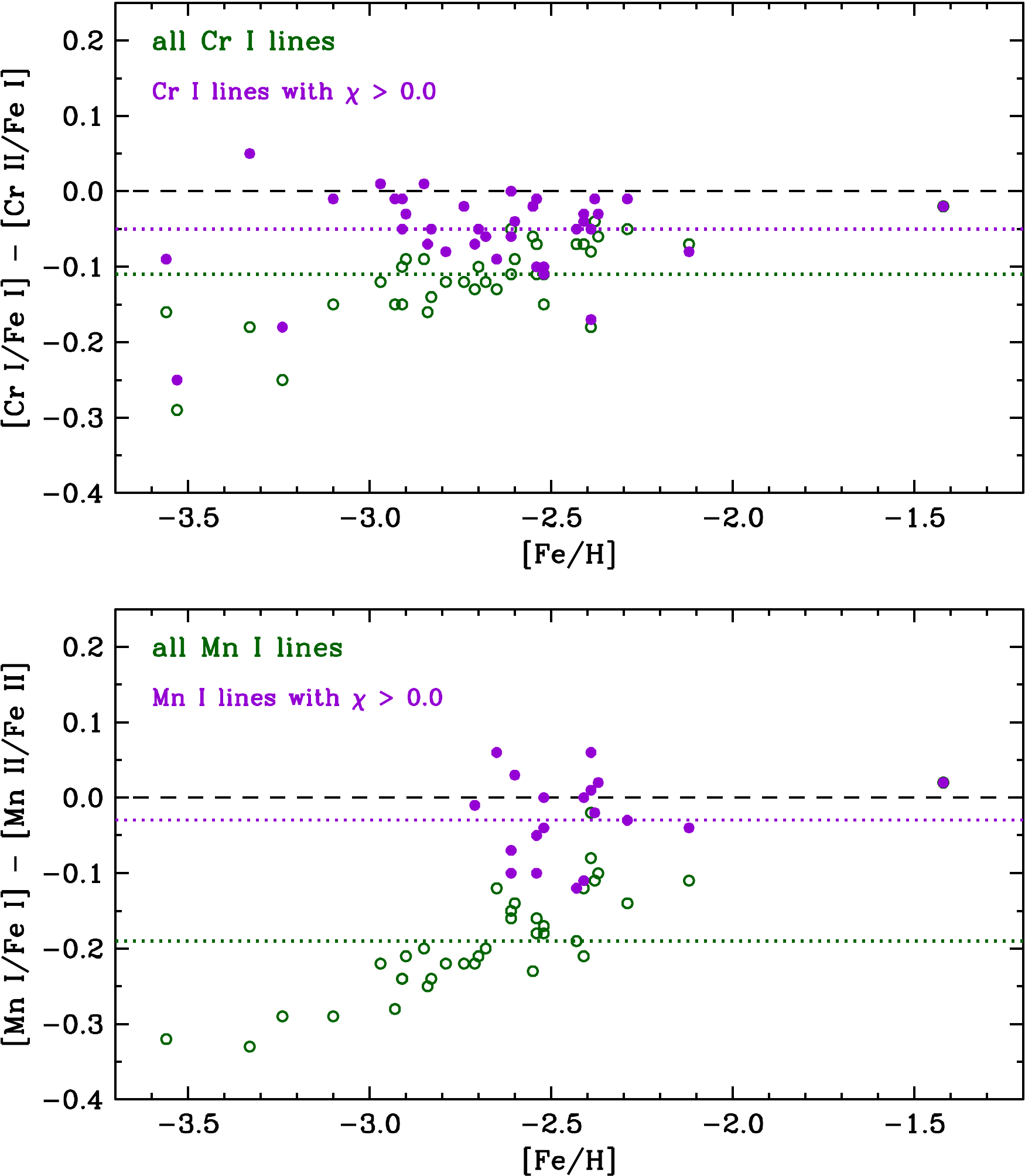}
\caption{
\label{fig5} \footnotesize
   Abundance differences between derived neutral and ionized abundances for Cr 
   (top panel) and Mn (bottom panel).
   As noted in the panel legends, the dark green open circles are for the 
   differences if all neutral lines are considered, while the purple filled 
   circles are for differences that include only excited-state lines,
   $\chi$~$>$~0.0~eV.
   The black dashed lines are for perfect agreement between the two species.
   The dotted lines indicate the mean differences including all stars.
}
\end{figure}

Abundance computations were straightforward applications of the model
atmospheres described in \S\ref{modelatm} with the model parameters
of Table~\ref{tab-models} to the measured $EW$s.
For even-Z elements the transitions were treated as single lines using the 
$abfind$ option in $MOOG$.
For odd-Z elements the line hfs components were accounted for in option 
$blends$.
Table~\ref{tab-abunds} contains all abundances for all stars in [X/Fe]
form, where these values have been computed with like ionization states, that
is neutral species compared to \species{Fe}{i} and ionized species compared
to \species{Fe}{ii}.
These abundance ratios are displayed as functions of metallicity in
Figures~\ref{fig3} and \ref{fig4}.
Inspection of these figures suggests that for most elements the star-to-star
[X/Fe] scatter is small at most metallicities, and mean [X/Fe]
values generally are constant over the whole metallicity range.
Furthermore, for the 5 elements beside Fe that have abundances derived 
from both neutral and ionized transitions (Ca, Ti, V, Cr, Mn) there is general
abundance agreement between the two species.
In Table~\ref{tab-specmean} we list these mean values for each
species computed for the whole stellar sample.  
We will use these figures and tables here to discuss our abundance derivations,
and reserve astrophysical interpretation for \S\ref{discussion}.

In the present work we did not attempt to apply NLTE corrections to our LTE
abundances; a detailed study on this topic should be conducted in the future.
The abundance derivations of all of these elements/species have been discussed
in our previous paper in this series, \eg, \cite{sneden16} and \cite{cowan20},
and the various lab/stellar papers.  
Here we comment on just a few items of special note.

\textbf{\species{Ca}{i} and \species{Ca}{ii}:}
This is the first large-sample application of the recently published
transition probabilities by \cite{denhartog21}.
The neutral and ionized species yield consistent Ca abundances:
$\langle$[\species{Ca}{i}/Fe] $-$ [\species{Ca}{ii}/Fe]$\rangle$ = 
0.02 ($\sigma$~=~0.05).\footnote{
The neutral/ionized species abundance differences quoted in this section
are of course for those stars with both species observed; thus the mean
differences will be similar to but not exactly equal to the whole-sample
species differences listed in Table~\ref{tab-specmean}.}
The available \species{Ca}{ii} lines for our work include 3 in the 
$\lambda$3170 and 2 in the $\lambda$3720 spectral ranges.
All of these lines are strong, and most have substantial blending issues.
In the higher-metallicity regime of our stellar sample, the \species{Ca}{ii}
line profiles become too strong and complex to yield reliable abundances
from our $EW$ analytical methods.
However, this species remains an important ionization equilibrium tool for
the lowest-metallicity main sequence turnoff stars.

\textbf{\species{V}{i}:} 
This species produces only very weak transitions in
optical-region spectra of our metal-poor main sequence turnoff stars.  
We were only able to detect \species{V}{i} lines in 7 out of our 37
program stars, all of them at the high-metallicity end of our sample.
For these stars, the mean abundance difference is
$\langle$[\species{V}{i}/Fe] $-$ [\species{V}{ii}/Fe]$\rangle$ = 
0.04 ($\sigma$~=~0.10).  
This general agreement between the two V species is encouraging but should 
not be considered as the final word on this point.

\textbf{\species{Cr}{i} and \species{Mn}{i}:}
In very low metallicity stars most reported Cr and Mn abundances are based
on the neutral resonance lines of \species{Cr}{i} 
$\lambda\lambda$4254,4274,4289 and \species{Mn}{i} 
$\lambda\lambda$4030,4033, 4034.
Unfortunately, it has been known for some time that these transitions 
yield elemental abundances that are significantly smaller
than those from weaker higher-excitation neutral-species lines and 
especially those from ionized-species transitions.

We called attention to this issue in \cite{sneden16} and \cite{cowan20},
and now with our large stellar sample we can confirm and extend this 
suggestion.
In Table~\ref{tab-abunds} we show two entries for \species{Cr}{i} and 
\species{Mn}{i} abundances in each star:  one for the mean abundance computed
with all transitions, and one with exclusion of the resonance lines.
In Figure~\ref{fig5} we compare neutral and ionized Cr and Mn abundances,
with \species{Cr}{i} values computed two ways: first with all measured 
transitions and then with exclusion of the $\chi$~=~0~eV resonance lines.
The differences are easily seen in the figure.

For Cr, the mean abundance difference using all neutral species lines
is $\langle$[\species{Cr}{i}/Fe] $-$ [\species{Cr}{ii}/Fe]$\rangle$ =
$-$0.11 ($\sigma$~=~0.06), but it reduces to $-$0.05 ($\sigma$~=~0.06
with exclusion of the neutral 0~eV lines.
For Mn the effect is larger:
$\langle$[\species{Mn}{i}/Fe] $-$ [\species{Mn}{ii}/Fe]$\rangle$ =
$-$0.19 ($\sigma$~=~0.08) with all lines included, but only
$-$0.03 ($\sigma$~=~0.06) when ignoring the 0~eV neutral species lines.
Among Fe-group elements, ionized species dominate by number in the 
(\teff,\logg) conditions of metal-poor main-sequence turnoff stars (see
Figure~2 and associated text in \citealt{sneden16}).
For almost all elements of this study the ion/neutral number density ratio in
line-forming atmospheric layers N$_{ion}$/N$_{neutral}$~$>$~50, usually
much greater.  
The only exception is Zn, which has the largest ionization
potential (I.P.~=~9.39~eV) of any Fe-group element.
Generally ionic transitions should yield more reliable abundances than those
of minority-species neutral transitions.
Happily, high-excitation \species{Cr}{i} and \species{Mn}{i} lines yield
abundances in good accord with those of the dominant ions.

Our purely observational result is in accord with NLTE abundance studies.
\cite{bergemann10a} studied neutral and ionized Cr species, finding only
small NLTE corrections for \species{Cr}{ii} but upward shifts approaching
0.3~dex in \species{Cr}{i}; values for individual transitions were not 
included in that work.
And as noted by \cite{cowan20}, the NLTE computations by \cite{bergemann08}
suggest increases of 0.2$-$0.5~dex should be applied to \species{Mn}{i}
abundances in very metal-poor main sequence stars, with the most severe
shifts needed for the 0~eV resonance lines.
More detailed line-by-line NLTE computations for especially these species
will be welcome, but for this paper we simply will neglect abundances
from the 0~eV lines in final \species{Cr}{i} and \species{Mn}{i}
abundances listed in Table~\ref{tab-models} and displayed in 
Figure~\ref{fig4}.
These transitions will be ignored for the rest of this paper.

\textbf{\species{Co}{i}:}
Large-sample Co abundance surveys almost always rely on \species{Co}{i}
transitions.
These studies, \eg, \cite{cayrel04}, \cite{barklem05}, \cite{roederer14} 
usually report supersolar [Co/Fe] values for low metallicity stars, 
[Fe/H]~$<$~$-$2.5.
Additionally, \cite{bergemann10} suggested that applying NLTE corrections for
\species{Co}{i} lines should increase LTE-based Co abundances by up to 
$\sim$0.4~dex, yielding final abundance ratios [Co/Fe]~$>$~0.5 at 
[Fe/H]~$\sim$~$-$3.
Very large relative Co abundances, derived from either LTE or NLTE analyses,
have always been difficult to match to Galactic chemical evolution predictions,
\eg, \cite{kobayashi06,kobayashi11b}.
The \cite{cowan20} study of 3 very metal-poor main sequence turnoff stars
with HST/STIS $UV$ spectra included more than 15 \species{Co}{ii} transitions
in each star.
Their results showed a sharp clash between neutral and ionized species:
$\langle$[\species{Co}{i}/Fe]$\rangle$~$\simeq$~0.37, but
$\langle$[\species{Co}{ii}/Fe]$\rangle$~$\simeq$~$-$0.13.
This suggested that the ionized Co species yields the correct elemental
abundance; see the \citeauthor{cowan20} paper for more detailed discussion.

Unfortunately our spectra do not contain detectable \species{Co}{ii} features.
Our analyses suggest only a mild mean overabundance, 
$\langle$[\species{Co}{i}/Fe]$\rangle$~=~0.21 (Table~\ref{tab-specmean},
but with a gradual increase toward decreasing metallicity (Figure~\ref{fig4}).
We urge caution in interpretation of our Co abundances.

\textbf{\species{Cu}{i}:}
Since these abundances are based mainly on the resonance lines in the
crowded $\lambda$3200 spectral region, we verified our $EW$ analyses with
full synthetic spectra.  
Our LTE result is $\langle$[\species{Cu}{i}/Fe]$\rangle$~=$-$0.71, nearly 
constant over the entire metallicity range.
However, multiple NLTE studies (\eg, \citealt{andrievsky18}, \citealt{shi18}) 
argue that the ground state electron population is significantly depleted
compared to LTE calculations, and suggest that the actual [\species{Cu}{i}/Fe]
values are $\sim$0.5~dex higher than those derived in LTE.
Abundances derived from \species{Cu}{II} lines in UV spectra from 
HST/STIS are higher by several tenths of a dex than those derived from 
\species{Cu}{I} lines, affirming the results of NLTE calculations 
\citep{korotin18,roederer18a}.
Like Cr and Mn, we report our LTE Cu abundances and urge comprehensive
NLTE studies of all Fe-group elements in the future.

\section{DISCUSSION\label{discussion}}

Elemental abundances in old low metallicity stars provide insight into the
early nucleosynthesis and star formation history in the  Galaxy. 
In particular the observed elements, Fe-peak along with Ca and Mg,  were 
synthesized in a previous generation (or generations) of stars. 
Such studies can also be made in globular clusters employing some of those 
elements (Sc, V and Zn) to probe early Galactic formation and evolution
\citep{minelli21}.

\begin{figure}
\epsscale{0.7}
\plotone{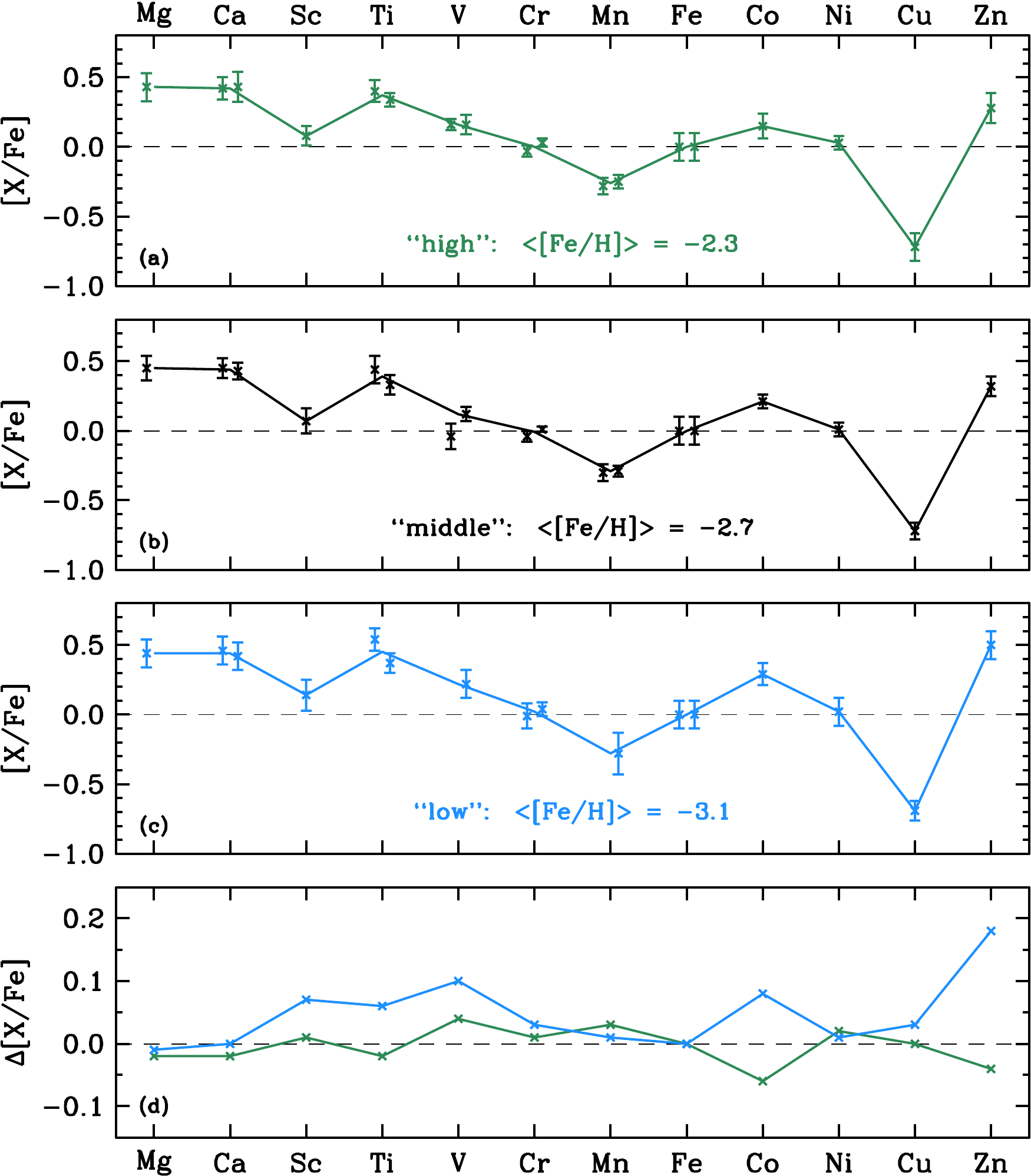}
\caption{
\label{fig6} \footnotesize
   Final species and elemental abundances for the low, middle,
   and high metallicity groups of our sample.  
   Panels (a), (b), and (c) show species abundance points 
   (Table~\ref{tab-specmean}) and elemental abundance lines 
   (Table~\ref{tab-elmean}) for the high, middle, and low abundance groups.
   For elements with abundances from two species, the points for neutral
   and ionized values are shifted by small amounts leftward and
   rightward of the element position, respectively.
   Panel (d) shows differences between the abundances of the high and middle
   metallicity groups (green line and points) and between those of the low
   and middle metallicity groups (blue line and points).
}
\end{figure}

Table~\ref{tab-elmean} summarizes our elemental abundance results.
Column~2 of the table contains the mean abundances for our entire 37 star 
sample, drawing from the species abundances of Table~\ref{tab-specmean}.  
For elements represented by only one species (Sc, Co, Ni, Cu, and Zn) the
elemental abundances are simply the species values.
For Ca, Ti, Cr, and  Mn the elemental abundances are averages of the neutral
and ionized species abundances.
And for V the ion abundance is exclusively used for the element because we 
had relatively few stars with neutral abundances for this element.
Table~\ref{tab-elmean} also lists in columns 3$-$5 the elemental 
abundance means for the 12 stars with lowest metallicities in our sample, 
the 12 stars with ``middle'' metallicities, and the 13 stars with highest
metallicities.
The mean metallicities for each group are listed in the column headers.

In Figure~\ref{fig6} we illustrate these abundance results.
In panels (a)$-$(c) we show the abundances for the 3 metallicity groups,
using data from Tables~\ref{tab-specmean} and \ref{tab-elmean}.
The dots with error bars are individual species abundances.
For those elements with both ionic and neutral measurements the abundance 
determinations agree well.
Inspection of these panels reveals overall abundance pattern agreement among 
the metallicity groups.
In particular, as earlier suggested by \cite{cowan20} with a sample of only 
four stars but including HST/STIS spectra below 3000~\AA, a distinct non-solar 
abundance pattern is evident:
\begin{itemize}
\item the traditional $\alpha$ elements\footnote{
Defined as the light elements whose major isotopes are multiples of
$\alpha$ nuclei. 
These include observable elements O, Mg, Si, S, and Ca.
Ti is often included with the $\alpha$'s because it is overabundant in 
metal-poor stars. 
However, its dominant isotope, \iso{48}{Ti}$_{22}$, is not an integer 
multiple of $\alpha$ particles.}
Mg and Ca are overabundant with respect to Fe by $\simeq$0.4~dex.
This confirms decades of previous work on these elements; our contribution
is the extensive use of \species{Ca}{ii} transitions to bolster our abundance
for this element.
\item the lightest Fe-group elements Sc, Ti, and V are overabundant with 
respect to Fe.
This issue was discussed by \cite{sneden16} and \cite{cowan20}; here we
add a large stellar sample to the discussion.
We will consider this in more detail below.
\item Mn is underabundant, but by less than 0.3~dex.
This is much smaller than the results presented in early abundance 
studies of metal-poor stars.
Neglect of the \species{Mn}{i} ground state lines and emphasis on 
\species{Mn}{ii} abundances, as discussed in \S\ref{abunds}, leads to this
mild Mn deficiency conclusion.
\item Co appears to be mildly overabundant, especially in the lowest
metallicity stars.
Caution should accompany interpretation of this result, as \cite{cowan20}
found no [Co/Fe] anomalies when studying UV \species{Co}{ii} lines in
3 stars with [Fe/H]~$\simeq$~$-$3.
The apparent overabundances from our optical-range \species{Co}{i} transitions
may be evidence of line formation issues rather than real abundance effects.
\item Zn is significantly overabundant in all 3 metallicity groups.
\end{itemize}
In panel (d) of Figure~\ref{fig6} we compare the elemental abundances of
the low and high metallicity groups to those of the middle group, to illustrate
trends with decreasing [Fe/H].
There is little difference between the middle and high metallicity groups. 
However, it is clear (see also Table~\ref{tab-elmean}) that abundance ratios
are different in the lowest metallicity group: several elements are
overabundant with respect to iron.
In particular, the Sc, Ti and V anomalies increase with decreasing metallicity.
[Co/Fe]  and [Zn/Fe] also show significant increases at lower metallicities.

\subsection{Elemental Abundance Correlations\label{correlations}}

\begin{figure}
\begin{center}
\includegraphics[angle=-90,width=2.4in]{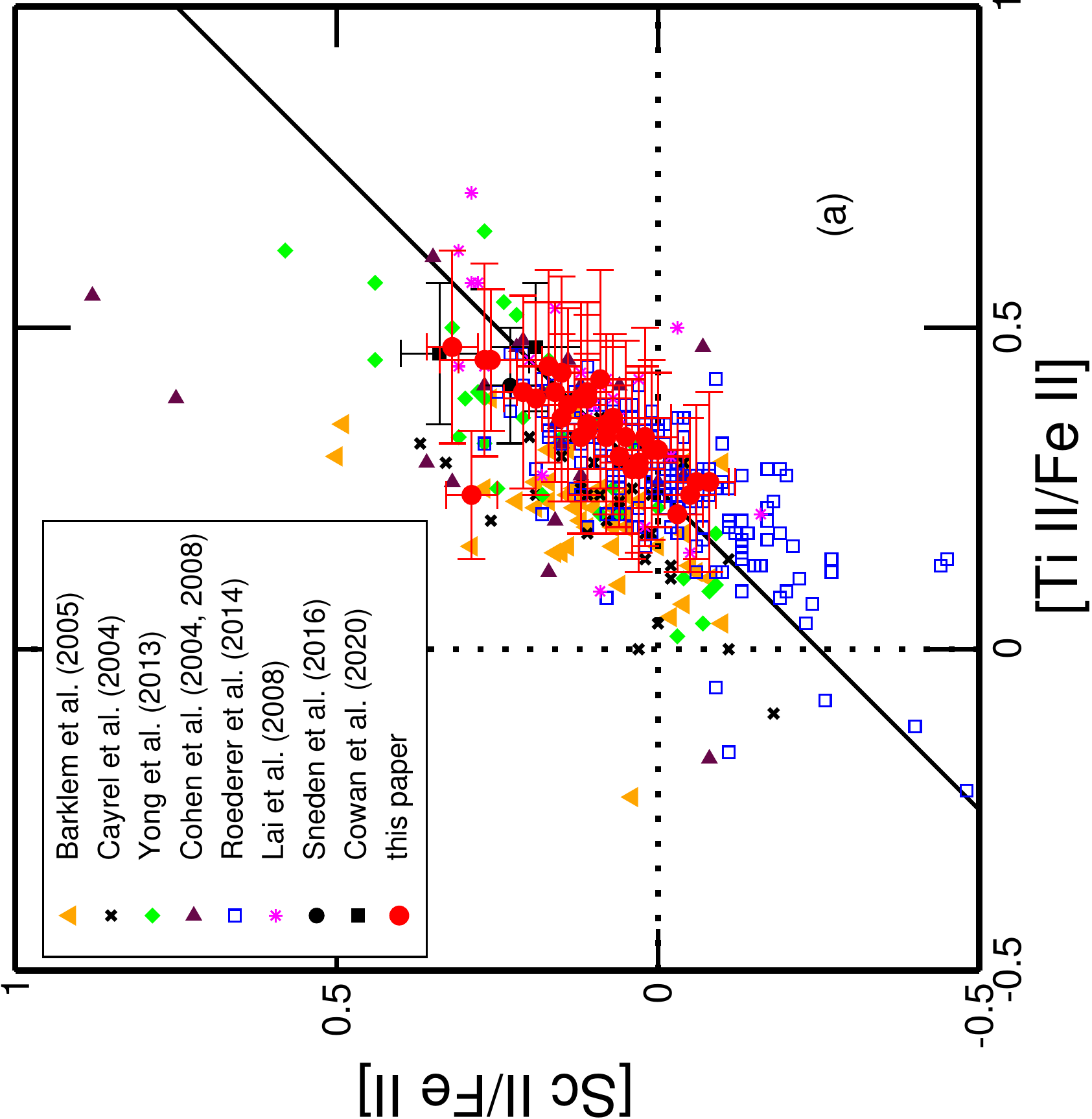}
\includegraphics[angle=-90,width=2.4in]{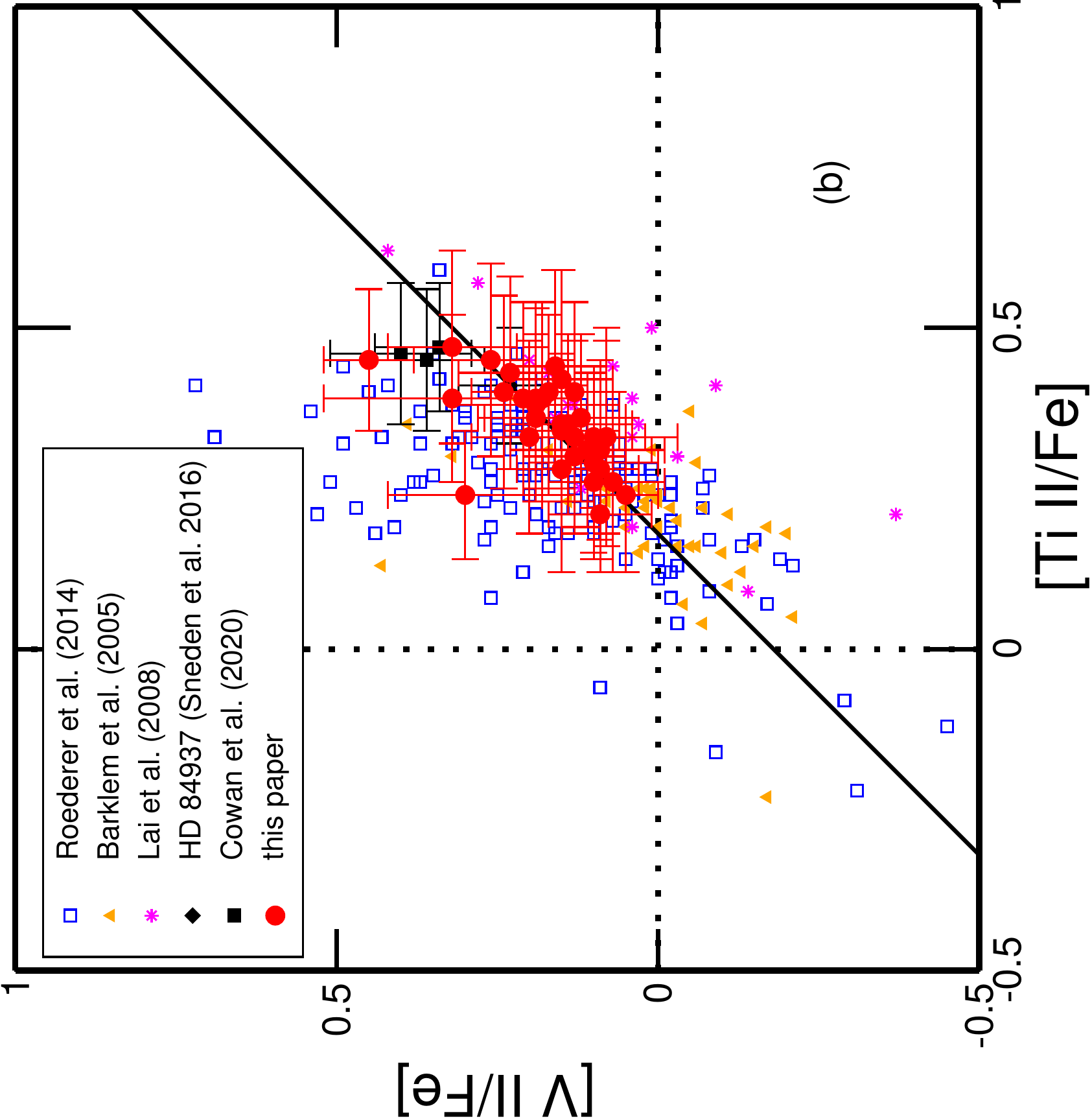}
\includegraphics[angle=-90,width=2.4in]{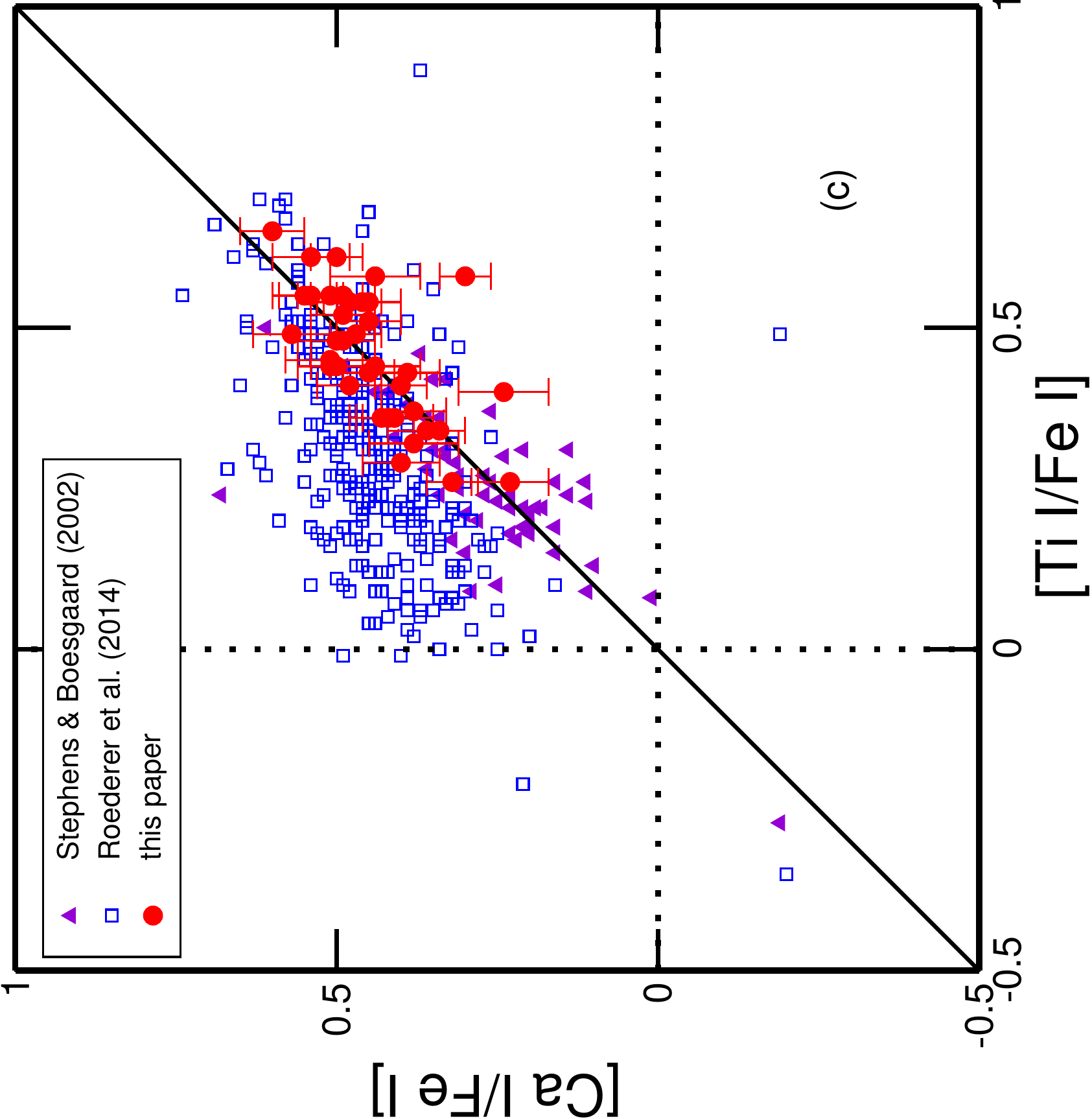}
\includegraphics[angle=-90,width=2.4in]{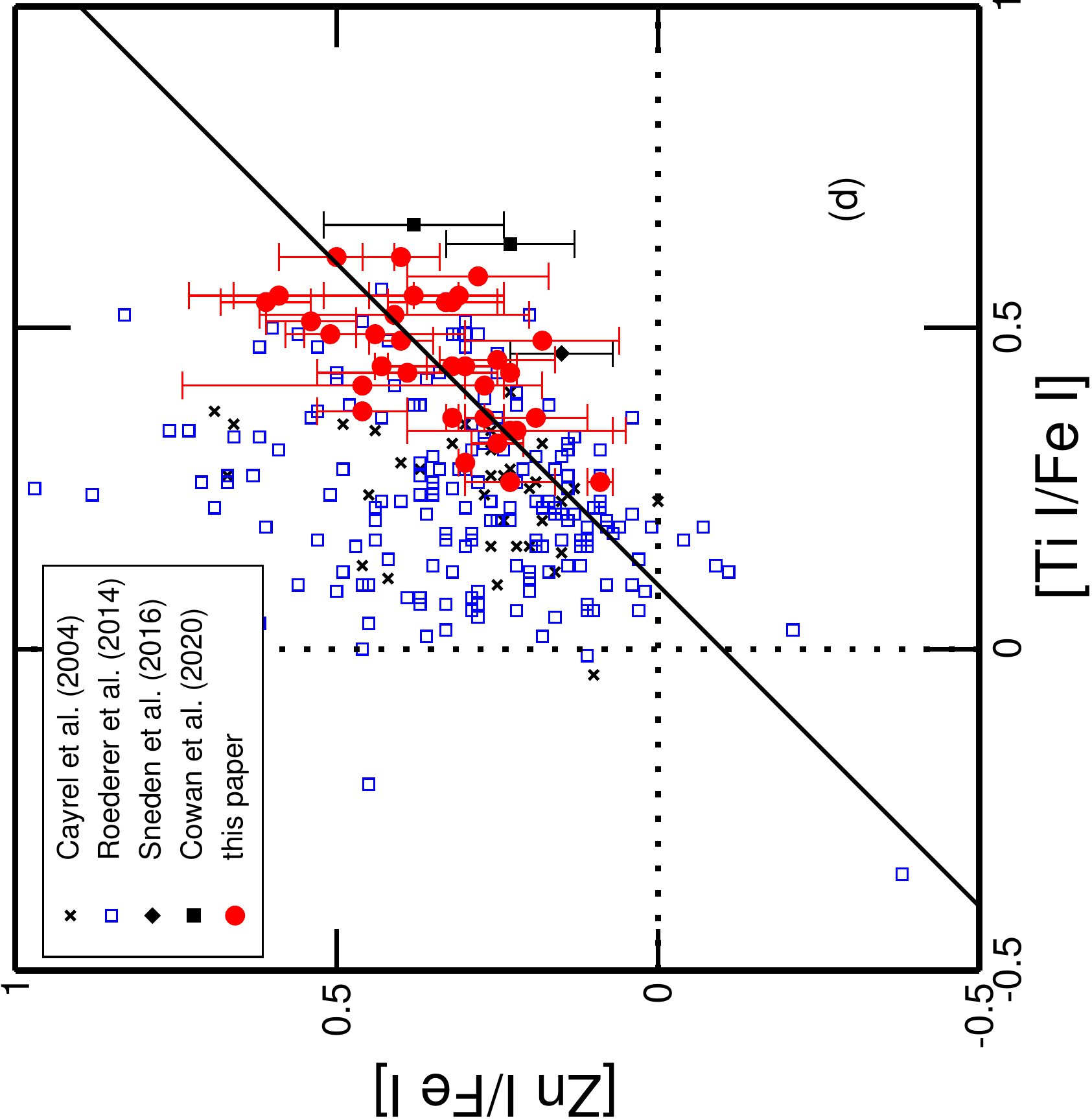}
\caption{
\label{fig7}\footnotesize
   Abundance ratios of 4 elements with respct to [Ti/Fe].
   These exhibit likely positive correlations.
   (a), \species{Sc}{ii} vs. \species{Ti}{ii};
   (b), \species{V}{ii} vs. \species{Ti}{ii};
   (c), \species{Ca}{i} vs. \species{Ti}{i}; and
   (d), \species{Zn}{i} vs. \species{Ti}{i}.
   Sources for the symbols are defined in the figure legends.
   \nocite{barklem05} \nocite{cayrel04} \nocite{yong13}
   \nocite{cohen04,cohen08} \nocite{roederer14} \nocite{lai08}
   \nocite{sneden16} \nocite{cowan20}
   The horizontal and vertical (dotted) lines denote the solar abundance ratios
   of each element.
   The 45$^{\circ}$ solid line represents perfect correlations between the
   abundance ratios.
}
\end{center}
\end{figure}

The similarity in the (non-solar) abundance patterns seen in 
Figure~\ref{fig6} suggests that the abundance ratios of some elements
are correlated.
Several such correlations are easily seen in Figure~\ref{fig7},
pointing to a possible common origin or nucleosynthesis history.
As in our past studies (\eg, \citealt{sneden16,cowan20}) 
we employ Ti as the comparison element. 
We have included earlier data sets that are identified in the figure legends.
These older abundances are only for comparison purposes and are not meant 
to be a comprehensive listing of all earlier studies.
The dotted horizontal and vertical lines indicate solar system values. 

\begin{figure}
\begin{center}
\includegraphics[angle=-90,width=2.4in]{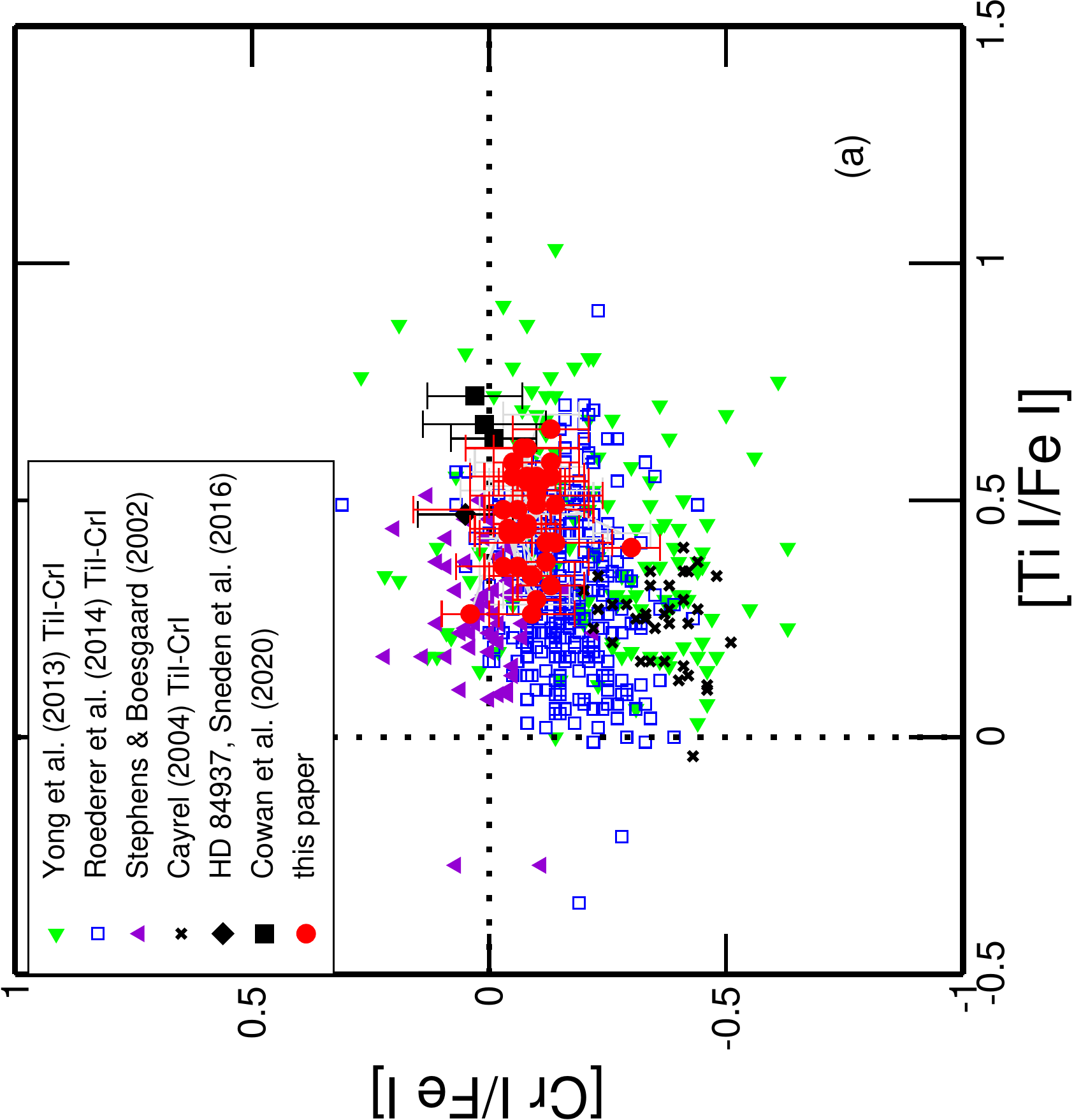}
\includegraphics[angle=-90,width=2.4in]{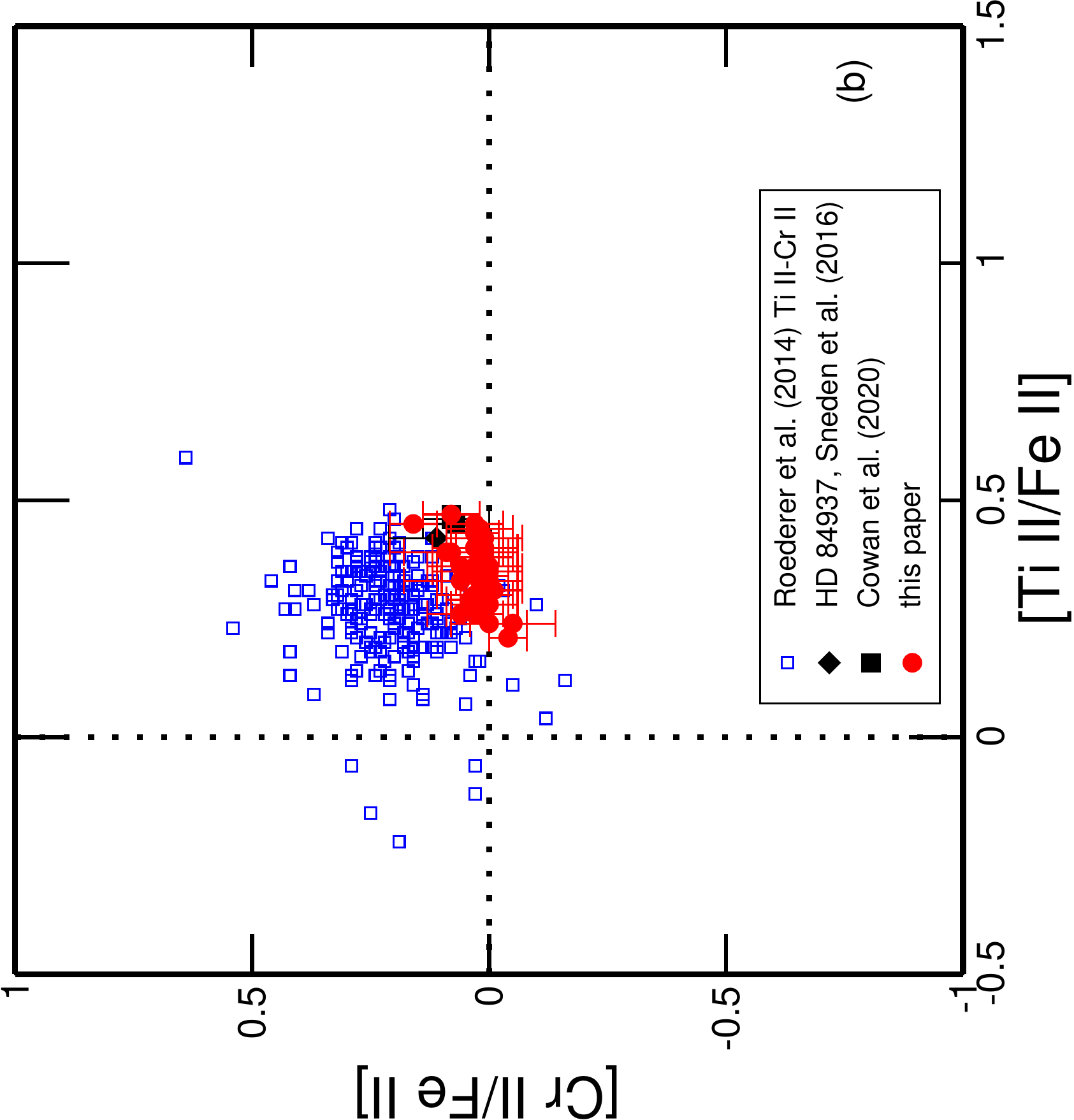}
\includegraphics[angle=-90,width=2.4in]{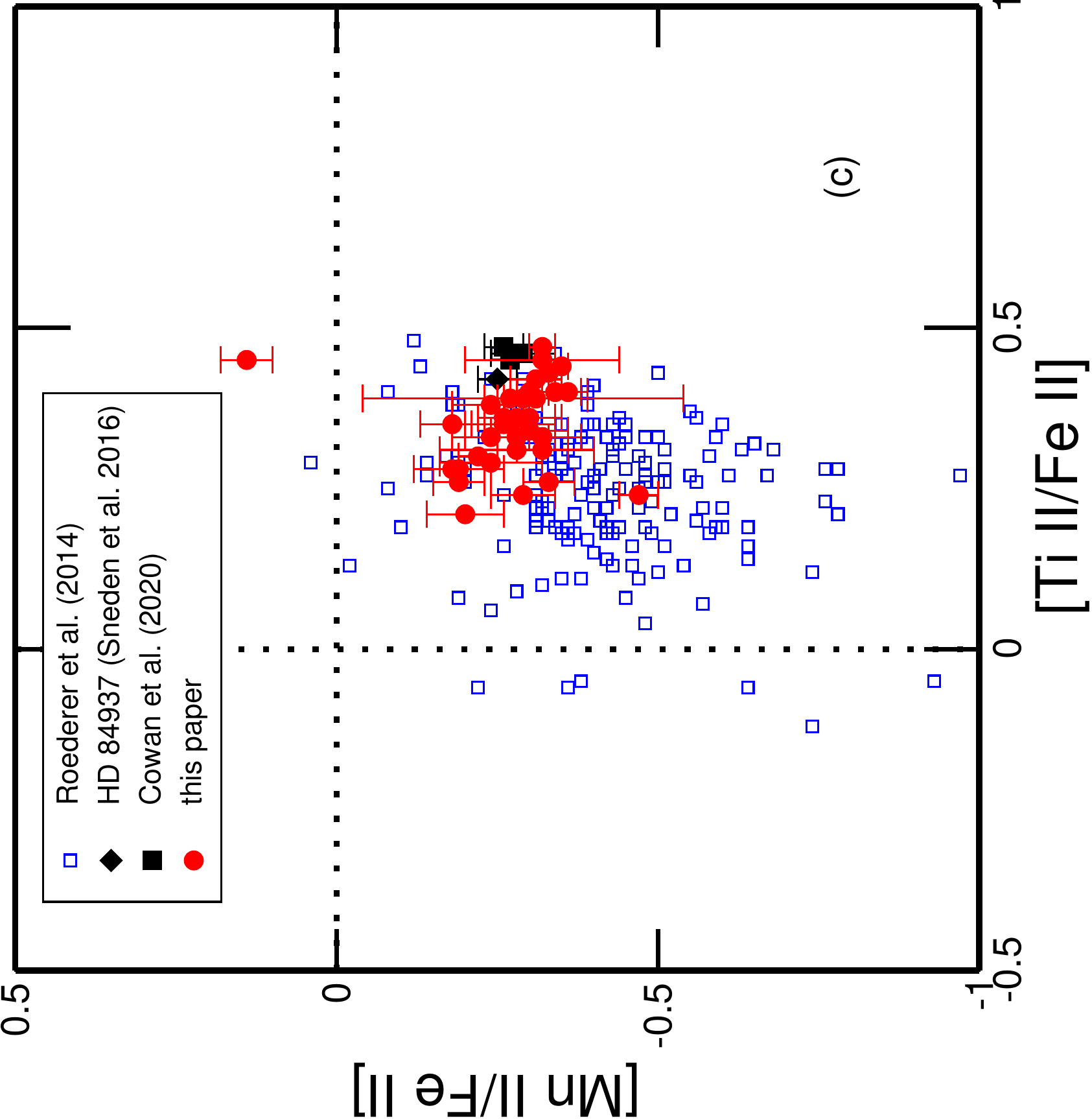}
\includegraphics[angle=-90,width=2.4in]{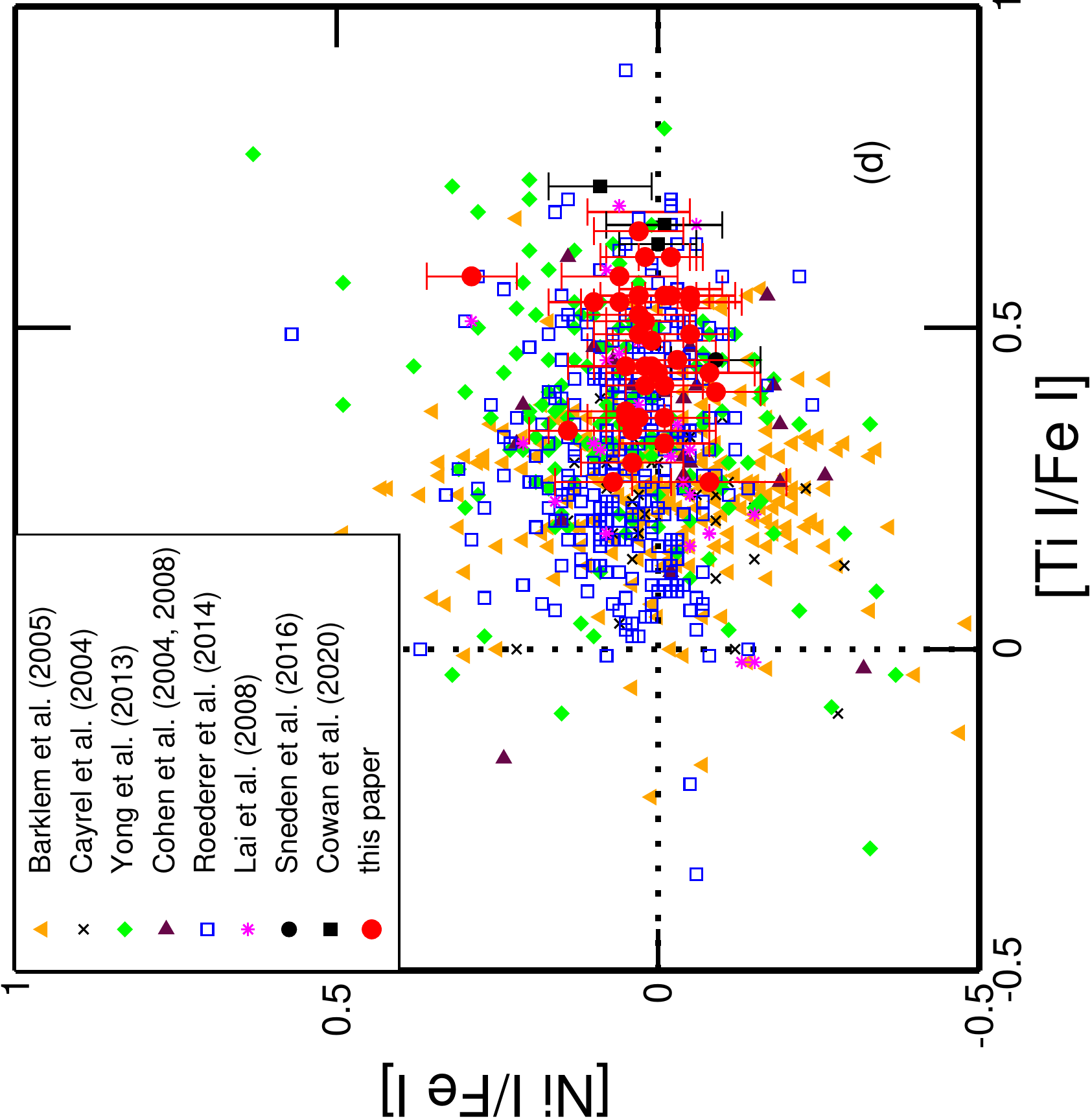}
\caption{
\label{fig8}\footnotesize
   Abundance comparisons indicating no correlations.
   (a) \species{Cr}{i} vs. \species{Ti}{i};
   (b) \species{Cr}{ii} vs. \species{Ti}{ii};
   (c) \species{Mn}{ii} vs. \species{Ti}{ii};
   (d) \species{Ni}{i} vs. \species{Ti}{i}.
   Symbols are defined in the panel legends and are consistent with
   those of Figure~\ref{fig7}.
}
\end{center}
\end{figure}

In panel (a) of Figure~\ref{fig7} we have plotted the values of 
\species{Sc}{ii} and \species{Ti}{ii} as red filled circles with error bars 
(line-scatter $\sigma$ values).
Inspection of these abundances reveals substantial correlation between
the two elements.
Other data sets generally support the trends in our data, although we have
not attempted to apply normalization corrections to align the other results
and ours.
To emphasize these correlations we have added a line at a 45$^{\circ}$ angle 
passing through the midpoint of our data.
In panels (b), (c), and (d) we have repeated these steps, although the
other elements have been studied in fewer previous papers.

We computed Pearson correlation coefficients for our data.
Defining $r$(X,Y)~$\equiv$ $r$([X/Fe],[Y/Fe]), we found 
$r$(\species{Sc}{ii},\species{Ti}{ii})~=~0.70,
$r$(\species{V}{ii},\species{Ti}{ii})~=~0.61,
$r$(\species{Ca}{i},\species{Ti}{i})~=~0.66, and
$r$(\species{Zn}{i},\species{Ti}{i})~=~0.59,
all positive moderate-to-high correlations ($|r|$~$>$~0.5).
By way of contrast the pure $\alpha$ element Mg shows no apparent abundance
correlation with Ti: $r$(\species{Mg}{i},\species{Ti}{i})~=~0.14.
The links between Ca, Sc, Ti, V, and Zn appear to be well established.
Note particularly that our new data, as well as that from \cite{roederer14}, 
suggest a correlation with increasing Ca tracking increasing Ti. 
This link has not been noted previously.
Our confidence in this correlation is increased by the determination of
abundances from \species{Ca}{ii} transitions in 25 stars in our sample of 37.
The neutral and ion abundances correlate well:
$r$([\species{Ca}{i}/Fe],\species{Ca}{ii}/Fe])~=~0.88.
Calcium is normally thought to be formed in explosive oxygen or (incomplete) 
silicon burning (\citealt{curtis19}); based upon this analysis, it may be 
formed in concert with Ti in early nucleosynthetic environments, preceding 
most halo star formation. 
Additional observational abundance data would help to confirm this correlation.

The correlations among [Sc/Fe], [V/Fe], and [Ti/Fe] have been noted 
previously (\eg, \citealt{sneden16}, \citealt{ou20}, \citealt{cowan20}),  
but the evidence is even clearer now with the addition of new precise 
abundance values in a larger sample of metal-poor stars. 
Ti and V are formed as a result of complete or incomplete Si burning, 
respectively (\citealt{curtis19}, \citealt{ebinger20}).
Although Ti synthesis is complicated and results from several processes 
in explosive environments (e.g., supernovae, SNe), our results suggest that 
at early Galactic times the synthesis sites for Fe-peak element production 
made an overabundance of Ti (on average by 0.3-0.5 dex).
While typical core-collapse SNe models have difficulty producing enhanced 
Ti at early Galactic times (see further comments in \S\ref{nuc})
higher explosion energies in hypernovae can produce larger amounts
of certain Fe-peak elements \citep{umeda02,kobayashi06}.
This might suggest more energetic hypernovae early in the Galaxy 
(see \citealt{kobayashi20}, and discussion in \citealt{sneden16},
\citealt{cowan20}).

We find a positive correlation of Zn with Ti, illustrated 
in panel (d) of Figure~\ref{fig7}.
\cite{li22} have identified moderate correlations in some of their data 
between Sc and the alpha elements, as well as between Ti and Zn.  
We strengthen these conclusions and extend them to additional elements.
We caution that there is more scatter in our Zn data than in other elemental
abundances and the comparison to the Zn abundances of \cite{roederer14} is 
not encouraging, but there appears to be a possible trend with higher [Zn/Fe] 
in stars being associated with higher [Ti/Fe] values. 
Lack of sufficient internally consistent high SNR spectra has previously 
prevented seeing such a possible correlation, and again, more 
data will be required to confirm such a correlation. 

We examined several other elemental abundance ratios to search for possible 
correlations that may fortify or negate the positive linkages among the 
elements discussed above.
Four examples are displayed in Figure~\ref{fig8}.
The computed correlation coefficients are
$r$(\species{Cr}{i},\species{Ti}{i})~=~0.23 (panel a);
$r$(\species{Cr}{ii},\species{Ti}{ii})~=~0.49 (panel b);
$r$(\species{Mn}{ii},\species{Ti}{ii})~=~0.02 (panel c);
$r$(\species{Ni}{i},\species{Ti}{i})~=~0.11 (panel d).
These suggest very low positive correlations between Cr and Ti abundances, 
and no apparent connections between Mn, Ni, and Ti abundances (low correlations
are here considered to be $|r|$~$<$~0.3).
These results suggest that the elements Cr, Mn and Ni are not formed in the 
same manner or nucleosynthesis site as that of Ti.
We also note that in panel (a) the derived [\species{Cr}{i}/\species{Fe}{i}] 
abundance ratios, computed without employing the \species{Cr}{i} resonance 
lines, is significantly higher than found in previous studies $-$ the new 
ratios are consistent with solar values; see the discussion in \S\ref{abunds}.
It is also apparent from panels (a) and (b) that the new 
[\species{Cr}{i}/\species{Fe}{i}] values (without resonance lines) are 
consistent with the [\species{Cr}{ii}/\species{Fe}{ii}] abundance ratios.
Inspection of Tables~\ref{tab-specmean} and \ref{tab-elmean} and discussion
in \S\ref{abunds} also indicate a similar result for neutral and ionized
Mn transitions.

\subsection{Galactic Abundance Trends\label{gce}}

\begin{figure}
\begin{center}
\includegraphics[angle=-90,width=2.4in]{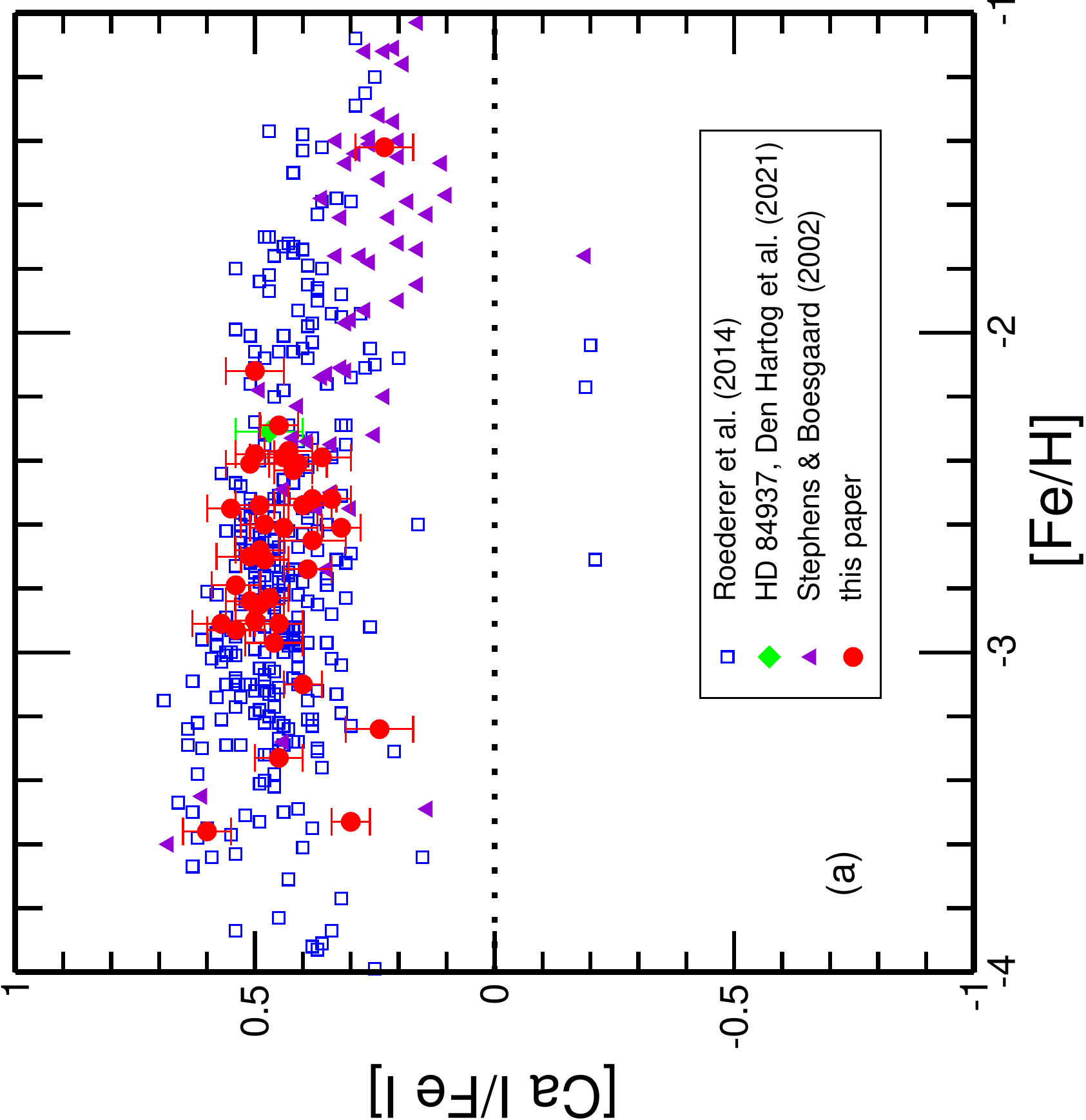}
\includegraphics[angle=-90,width=2.4in]{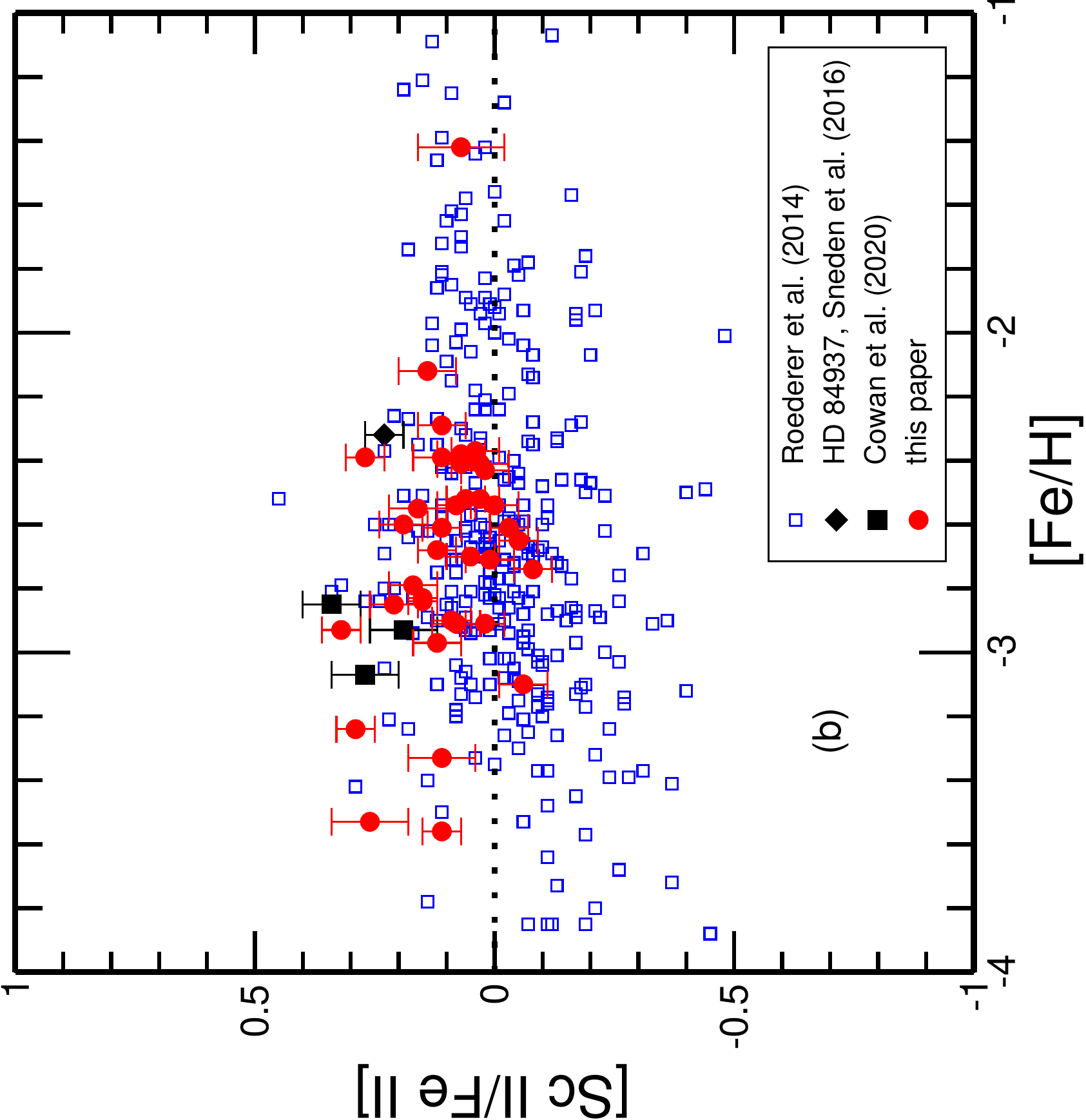}
\includegraphics[angle=-90,width=2.4in]{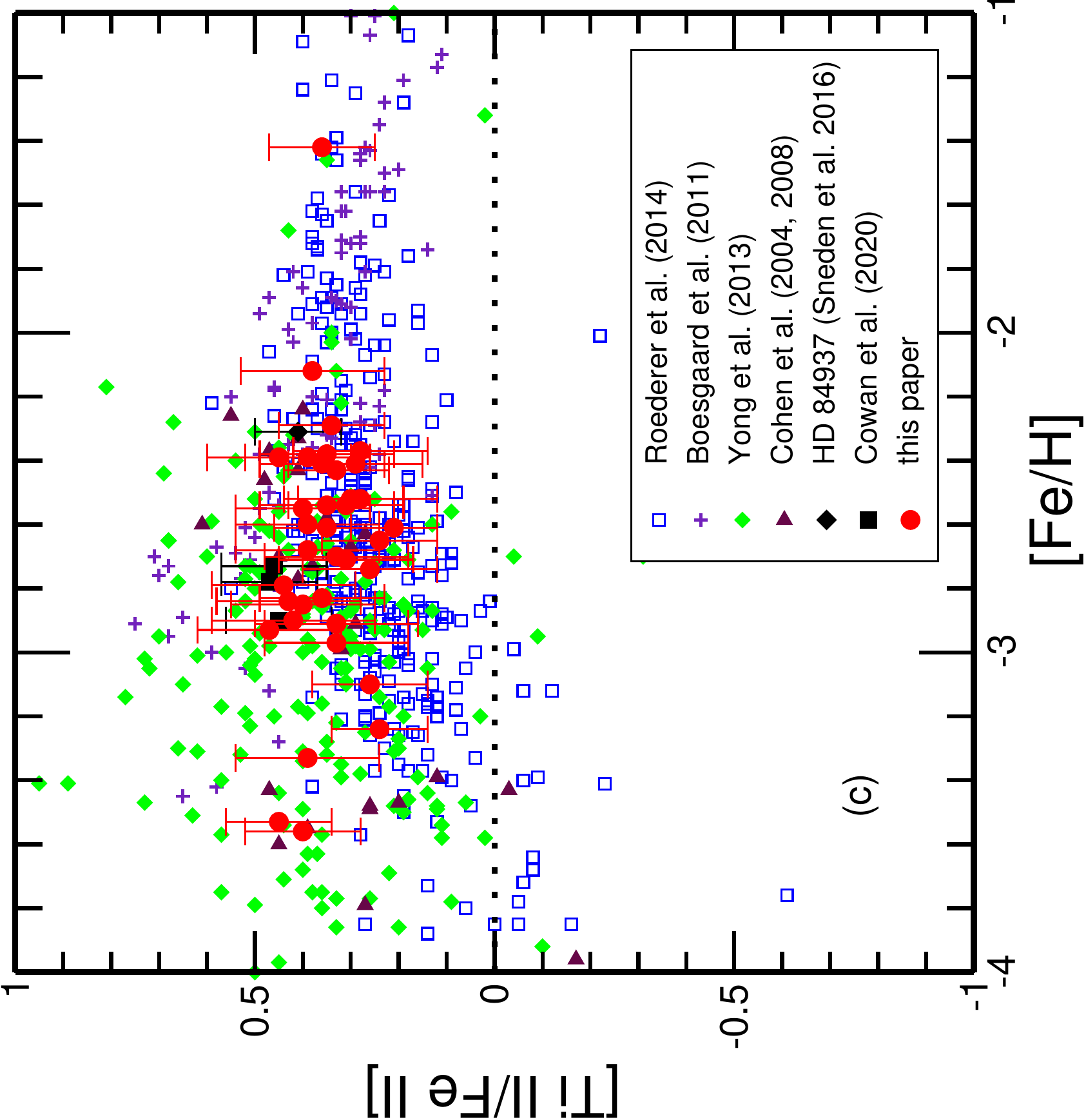}
\includegraphics[angle=-90,width=2.4in]{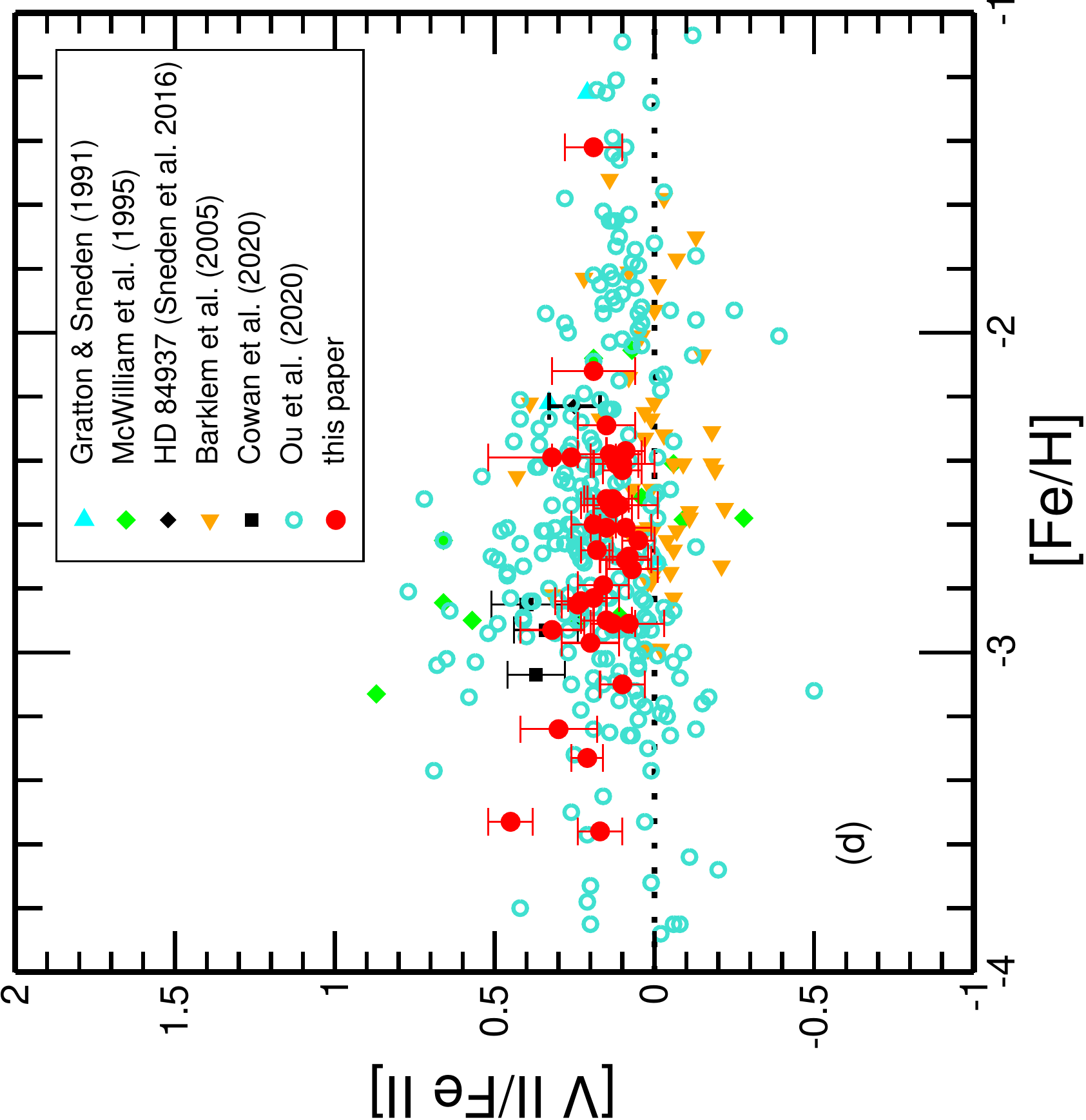}
\caption{
\label{fig9}\footnotesize
   Abundance ratios of various elements versus [Fe/H] metallicity.
   The symbols are defined in the figure legend and are consistent with the
   symbols in Figures~\ref{fig7} and \ref{fig8}.
   The horizontal (dotted) line denotes the solar abundance ratios
   of each element.
   \nocite{roederer14} \nocite{denhartog21} \nocite{stephens02} 
   \nocite{sneden16} \nocite{cowan20} \nocite{boesgaard11} \nocite{yong13} 
   \nocite{cohen04,cohen08} \nocite{gratton91} \nocite{mcwilliam95}
   \nocite{barklem05} \nocite{ou20}  
}
\end{center}
\end{figure}

\begin{figure}
\begin{center}
\includegraphics[angle=-90,width=2.4in]{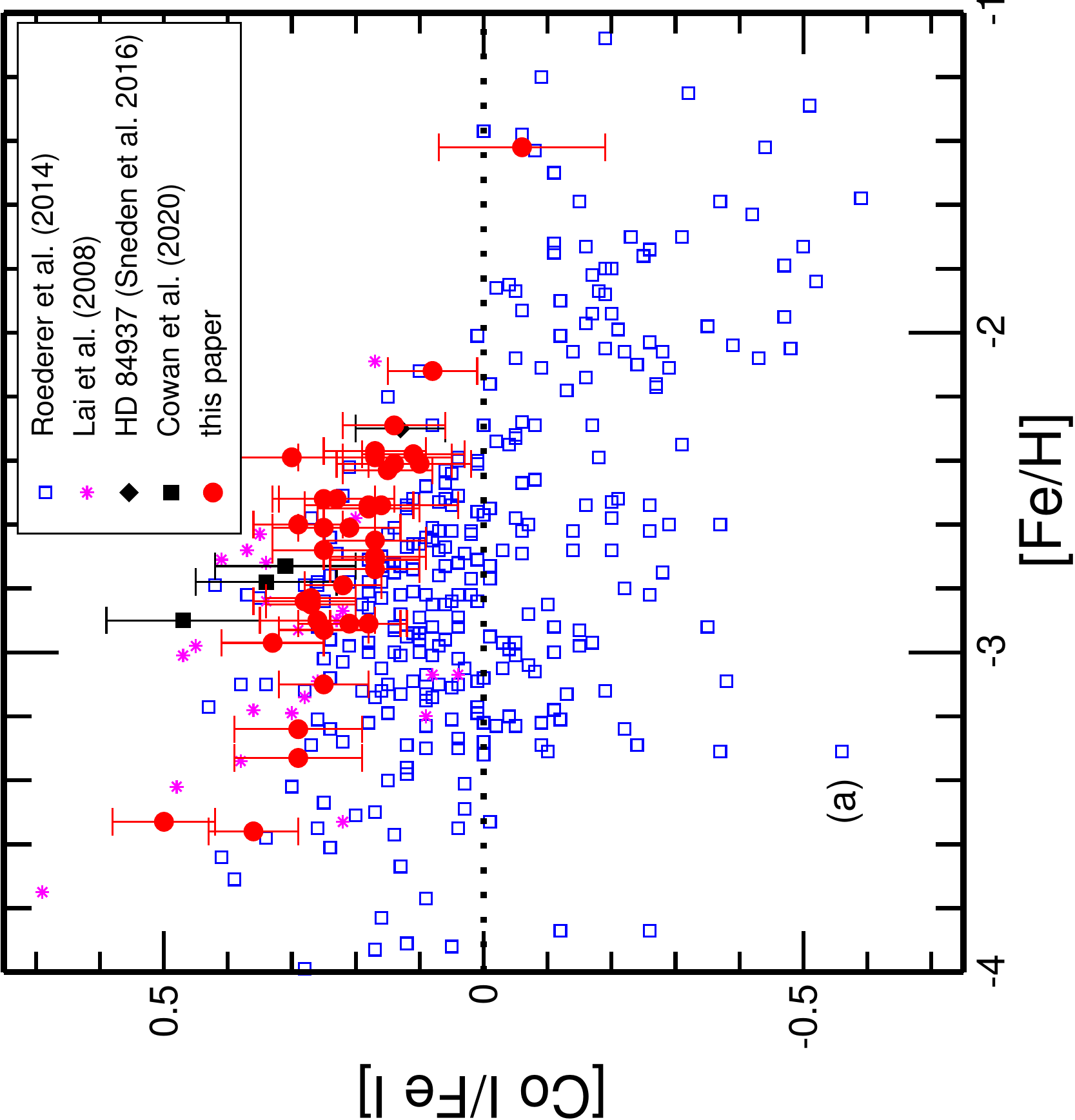}
\includegraphics[angle=-90,width=2.4in]{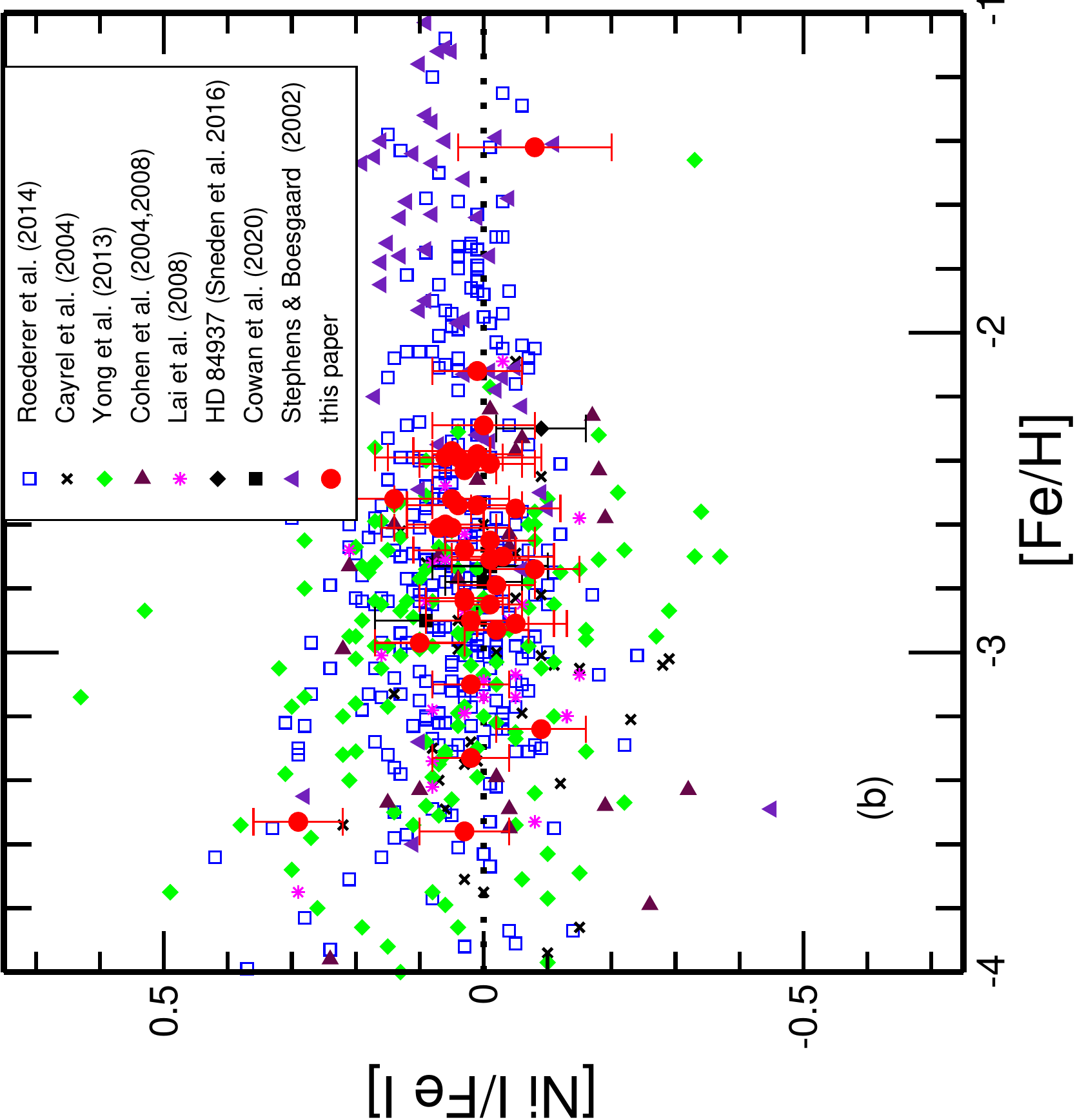}
\includegraphics[angle=-90,width=2.4in]{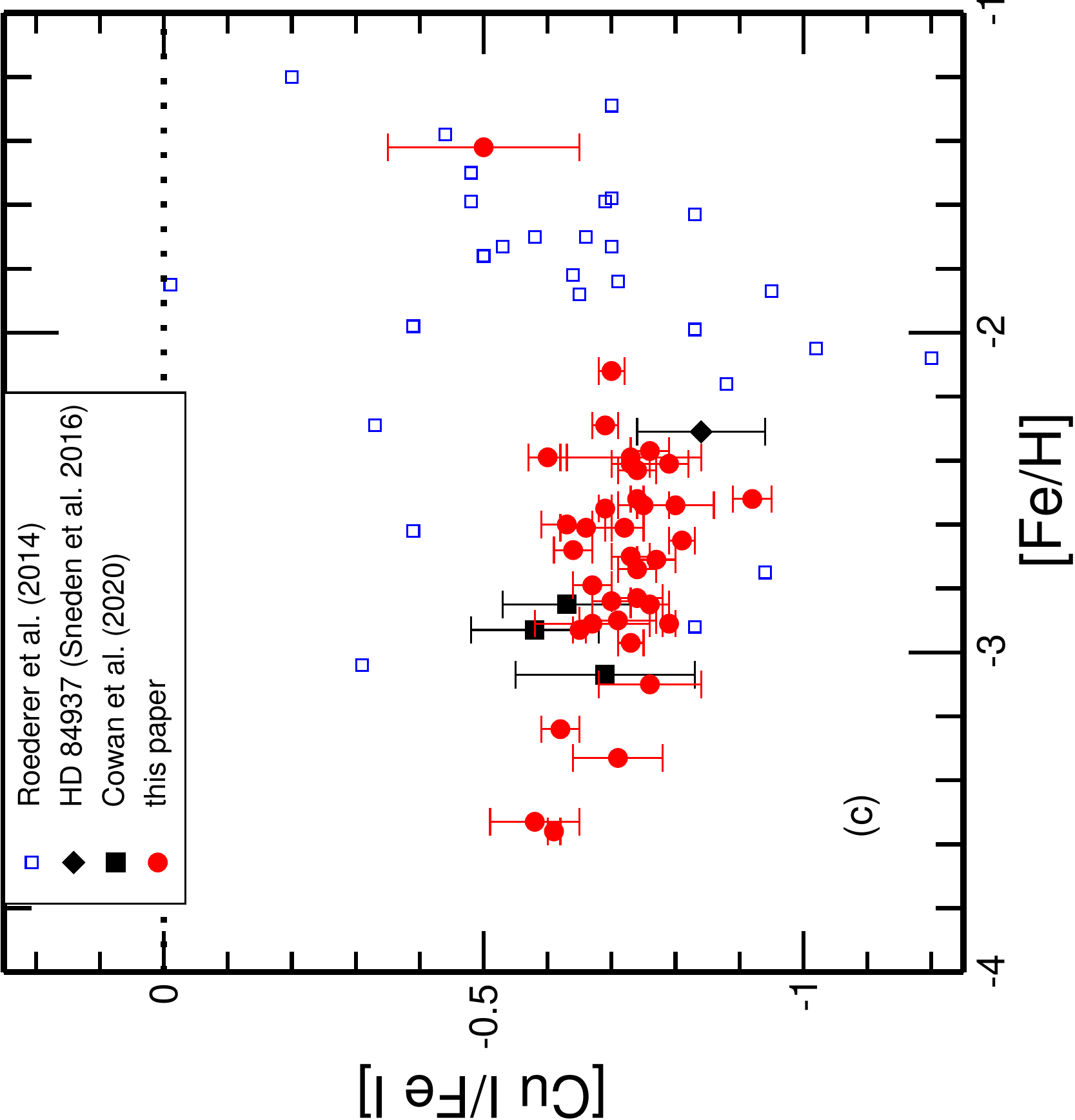}
\includegraphics[angle=-90,width=2.4in]{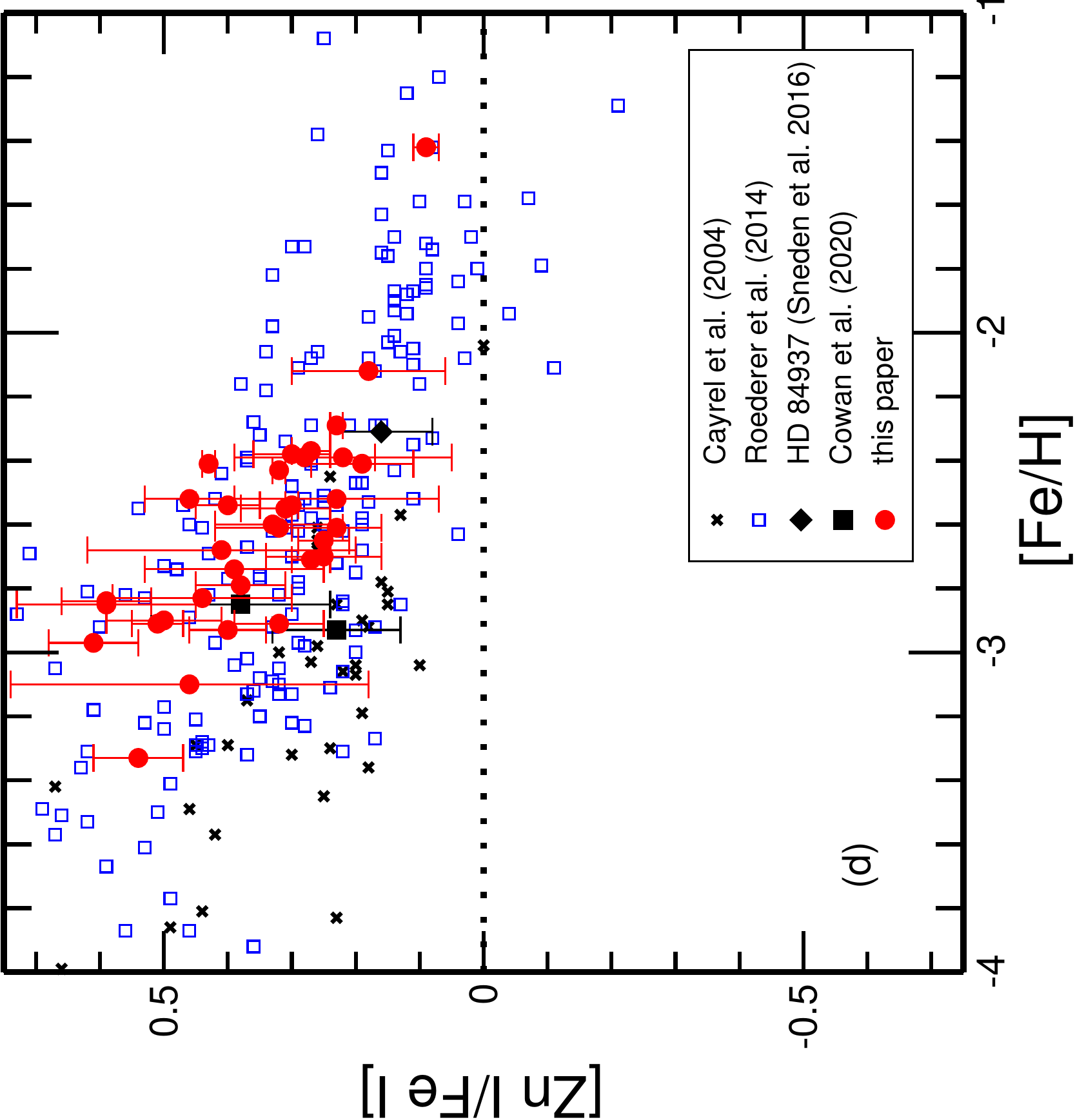}
\caption{
\label{fig10}\footnotesize
   (a): Abundance ratios of various elements versus [Fe/H].  
   See the figure legend for the data sources.
   \nocite{roederer14} \nocite{boesgaard11} \nocite{yong13} 
   \nocite{cohen04,cohen08} \nocite{sneden16} \nocite{cowan20}
   The error bars here are simply the line-scatter $\sigma$ values for each 
   star.
   The horizontal (dotted) lines in each panel denote the solar abundance 
   ratio.
}
\end{center}
\end{figure}

In Figures~\ref{fig9} and \ref{fig10} we illustrate abundance trends as 
functions of metallicity for the elements in our new survey.
Here we comment on them individually.

\textbf{Calcium:} 
Our [Ca/Fe] trend (Figure~\ref{fig9} panel a) is consistent with earlier 
results from \cite{stephens02}, \cite{roederer14}, and \cite{denhartog21},
all plotted in this panel.
These data show values above solar for a wide metallicity range. 
The downward trend in [Ca/Fe] at higher metallicity has long been interpreted
as due to increased production of iron from Type Ia SNe. 
Below metallicity [Fe/H]~$\sim$~$-$2 the Ca overabundance qualitatively appears
to plateau at [Ca/Fe]~$\sim$~0.5 if all data sets in Figure~\ref{fig9} are
included.
There may be a slight increase in Ca overabundance with decreasing 
metallicity in our stellar sample, but it is not statistically meaningful:
$r$(\species{Ca}{i},[Fe/H])~=$~$0.21, consistent with essentially no trend
with metallicity.
Future studies of [Ca/Fe] at the lowest metallicities are required to
pursue this point.

\textbf{Scandium:} 
Panel (b) of Figure~\ref{fig9} shows relatively large star-to-star scatter 
in our [\species{Sc}{ii}/\species{Fe}{ii}] ratios at a given [Fe/H] 
metallicity, and generally higher abundance ratios than those
reported by \cite{roederer14} at the lowest metallicities.
Our data suggest no significant metallicity trend for Sc.

\textbf{Titanium:} 
There is little indication of Ti overabundance variations with metallicity
in our sample, as illustrated in Figure~\ref{fig9} panel (c), and
as computed in a correlation coefficient:  $r$(\species{Ti}{ii},[Fe/H])~=~0.12.
The [Ti/H] vs [Fe/H] correlation
in Figure 15 of B11 show a tight
correlation between Ti and Fe abundances over almost 4 orders of magnitude in
[Fe/H] in their 117 stars.  
They found a slope of 0.86~$\pm$~0.01.
Results from other studies shown in Figure~\ref{fig9} panel (c) appear to 
be divergent at lowest metallicities; this should be pursued in
future studies.
At metallicities [Fe/H]~$\lesssim$~$-$2.5 the star-to-star scatters increase
within individual surveys.
The disagreements in scale among the surveys do not inspire confidence in 
the published [Ti/Fe] abundance ratios at lowest metallicities.
Resolution of this issue is beyond the scope of our study.
  
\textbf{Vanadium:} 
The [\species{V}{ii}/Fe] relative abundances shown in panel (d) of 
Figure~\ref{fig9} have little change with metallicity. 
$r$(\species{V}{ii},[Fe/H])~=~$-$0.30, at the low edge of a possible trend
with decreasing [Fe/H].

\textbf{Cobalt:} 
In panel (a) of Figure~\ref{fig10} we correlate 
[\species{Co}{i}/\species{Fe}{i}] with metallicity. 
The trend of rising Co abundance ratios at low [Fe/H] is easy to see and 
strongly backed statistically: $r$(\species{Co}{i},[Fe/H])~=~$-$0.84.
The apparent Co overabundances in the most metal-deficient stars was
not an artifact of weak \species{Co}{i} line measurement errors.
$EW$s for many \species{Co}{i} transitions were easily measured even in our
lowest metallicity stars.
Most lines detectable in these stars have wavelengths in the 3400$-$3600~\AA\
spectral domain.
But the original B11 requirement to have reasonable SNR down to 3100~\AA\ 
ensured very good SNR values in this part of the near-UV range.
Measurement of \species{Co}{i} lines with log~$RW$~$\gtrsim$~$-$5.8 was 
not difficult.

The increasing [Co/Fe] trend with decreasing metallicity, shared by 
other surveys included in panel (a) of Figure~\ref{fig10}, has been known 
since the survey of very metal-poor giants by \cite{mcwilliam95}; see their 
Figure~11.
However, as noted in \S\ref{abunds} and discussed in detail by \cite{cowan20},
this increase in [\species{Co}{i}/Fe] is not shared by [\species{Co}{ii}/Fe]
values.
Unfortunately, the ion abundances are only available via HST/STIS spectra,
inaccessible in this study and not likely to be increased in the foreseeable
future.
An NLTE investigation by \cite{bergemann10} suggested that LTE
[\species{Co}{i}/Fe] values should be $increased$ by as much as +0.4~dex 
very low metallicities, which would significantly increase the neutral/ion Co
abundance clash.
We regard the Co abundance problem as unsolved at this point, and urge
additional NLTE investigations.

\textbf{Nickel:} 
Inspection of panel (b) of Figure~\ref{fig10} immediately suggests no evolution
of [\species{Ni}{i}/Fe] values over the entire $-$4~$<$~[Fe/H]~$<$~$-$1 
metallicity range in any of the many studies of this element.
Our correlation coefficient confirms the qualitative assessment:
$r$(\species{Ni}{i},[Fe/H])~=~$-$0.01, excluding the one very discordant high
value for G~275-4 ([Fe/H]~=~$-$3.53).
The Ni abundance is well-determined in G~275-4, but it appears to be 
out of the mainstream of values for this element at low metallicity.
However only two stars of our sample have [Fe/H]~$\lesssim$~$-$3.3; general
conclusions on the Ni abundance in G~275-4 await a larger-sample study of
similar stars.

\begin{figure}
\epsscale{0.7}
\plotone{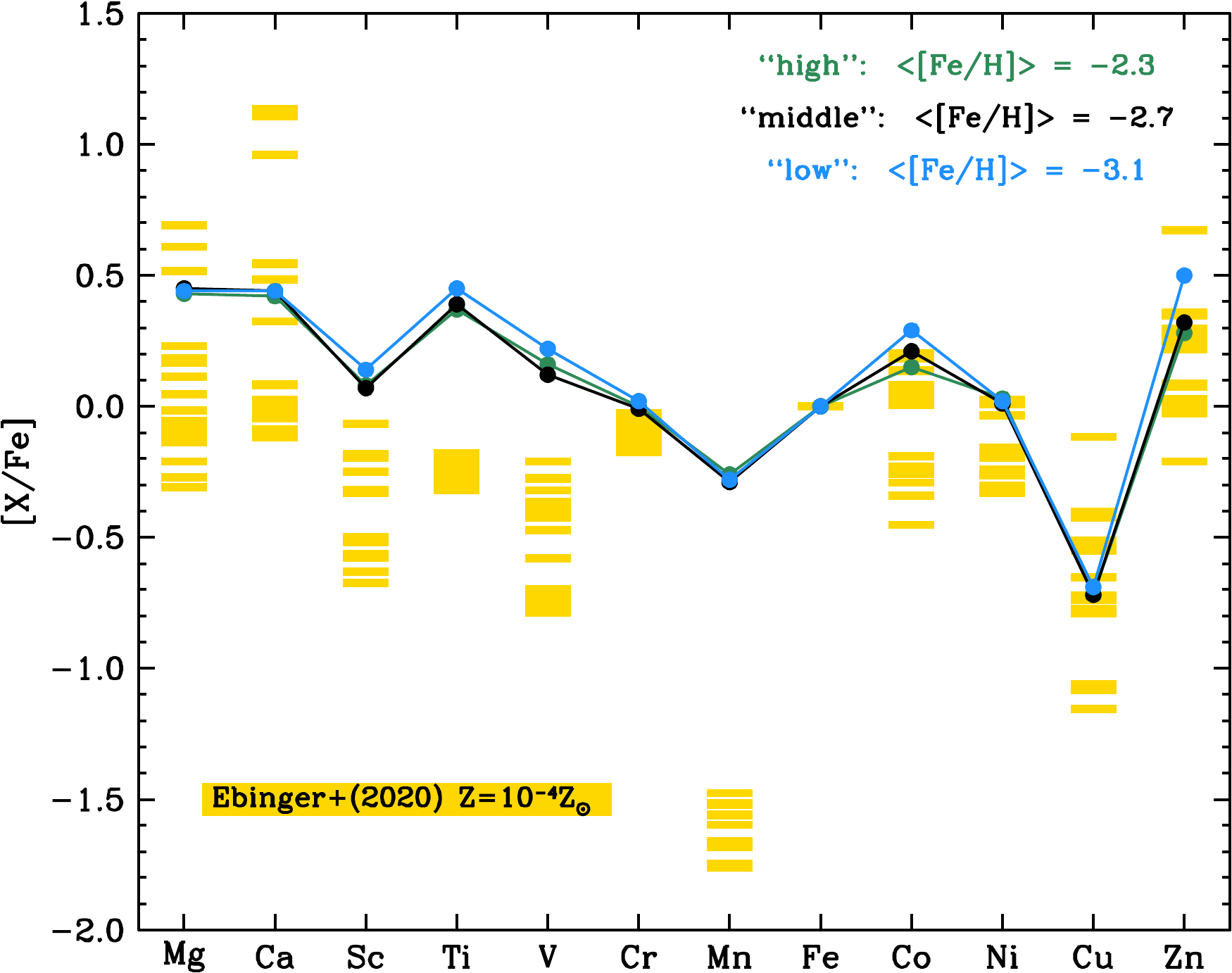}
\caption{
\label{fig11} \footnotesize
   Supernovae nucleosynthesis predictions from \cite{ebinger20} for 
   low-metallicity stars compared to average abundance determinations of 
   the stars from this paper (Table~\ref{tab-elmean}).
}
\end{figure}

\textbf{Copper:} 
Our [\species{Cu}{i}/Fe] values are based on LTE analyses, and as discussed
in \S\ref{abunds} they are likely to be deficient by $\sim$0.5~dex compared
to values derived in NLTE.
Thus our mean value, $\langle$[\species{Cu}{i}/Fe]$\rangle$~=~$-$0.71 
($\sigma$~=~0.08), might turn out to be $\sim -$0.2.
Our Cu abundances appear to be relatively constant.  
Note the lack of many literature sources on Cu in very low-metallicity stars.
There are few \species{Cu}{i} lines in optical wavelength regions redward of
4000~\AA.
The strongest line at 5105.5~\AA\ is too weak to be reliably detected in any
of our stars with [Fe/H]~$<$~$-$2.
Our Cu abundances are derived from the resonance lines at $\lambda$3247,3273.
Most abundance surveys of metal-poor stars do not extend blueward to
these wavelengths.

\textbf{Zinc:} 
Abundance evolution of Zn with metallicity is shown in panel (d) of 
Figure~\ref{fig10}.
We confirm increasing [\species{Zn}{i}/Fe] abundance with decreasing [Fe/H]
first identified in the large-sample survey of metal-poor red giants by
\cite{cayrel04}.
The correlation coefficient for our data agrees with the visual impression:
$r$(\species{Zn}{i},[Fe/H])~=~$-$0.74 .
The lines of \species{Zn}{i} at $\lambda$4722,4810 are often very weak in
our spectra, but our analyses by both EW and synthetic spectra agree well.
Seeing the same trend in both main sequence and red giant stars strengthens
the conclusion that this is a genuine abundance result, not an analytical
artifact.
While there is star-to-star scatter in our data and in literature results, 
it is clear that relative [Zn/Fe] production increases with decreasing 
metallicity.

\subsection{Nucleosynthesis  Origins\label{nuc}}

We have shown from a large sample of 37 very metal-poor main sequence turnoff
stars that high correlations exist between the elements Ca, Sc, Ti, V  and Zn,
suggesting they were formed in concert near the beginning of our Galaxy.
These elements are synthesized in several processes; see \cite{curtis19} 
and discussion in \cite{sneden16}, \cite{cowan20}. 
However, the observed correlations might suggest that one specific 
process (perhaps complete or incomplete silicon burning) might be the 
dominant nucleosynthesis path for these particular elements 
in early Galactic SNe.

We compare our measured abundance values with recent SNe calculations from 
\cite{ebinger20} in Figure~\ref{fig11}.  
They show nucleosynthesis yields from core-collapse SNe for solar and 
low-metallicity ([Fe/H] = $-$4) cases; the latter values might be typical for 
the progenitors to the stars of our survey.
We have illustrated a range of their model predictions (indicated by 
vertical, often overlapping horizontal bars) for each element from Mg-Zn.
It is seen that they correctly predict values for the low-metallicity stars
for many of our observed elements, particularly the heavier ones (Z~$\geq$~26).
It is noteworthy that they are able to reproduce our observed rises in
[Co/Fe] and [Zn/Fe] for all stars in the sample.  
The exceptions are Sc, Ti, V, and Mn, where the observed abundances [X/Fe] are 
larger than model predictions. 
These overabundances, particularly for Ti, have been well-known for some time.
It is interesting that these divergences between calculations and observations
occur for elements below iron (Z~$<$~26). 
This, in turn, might suggest some differences (i.e., decreases) in iron 
production (with respect to elements such as Ti) at these very low 
metallicities, and perhaps for the earliest Galactic SNe (while iron 
production in the Galaxy today is primarily from Type Ia SNe, core-collapse 
SNe were responsible for the earliest Galactic iron synthesis.) 
We also note new model predictions for Mg and Ca from \cite{ebinger20} also 
replicate our abundance determinations for those elements in the sample stars.
In the future it will be important to determine if there are rises in the 
abundances of those two elements at the lowest metallicities. 
Our new  results $-$ at an elemental abundance accuracy of 0.1 dex $-$ could  
provide constraints, not only at low metallicities but also at high 
metallicities, for SNe models early in Galactic history.

\subsection{Trends with Kinematics\label{kinemat}}

\begin{figure}
\begin{center}
\epsscale{0.9}
\plotone{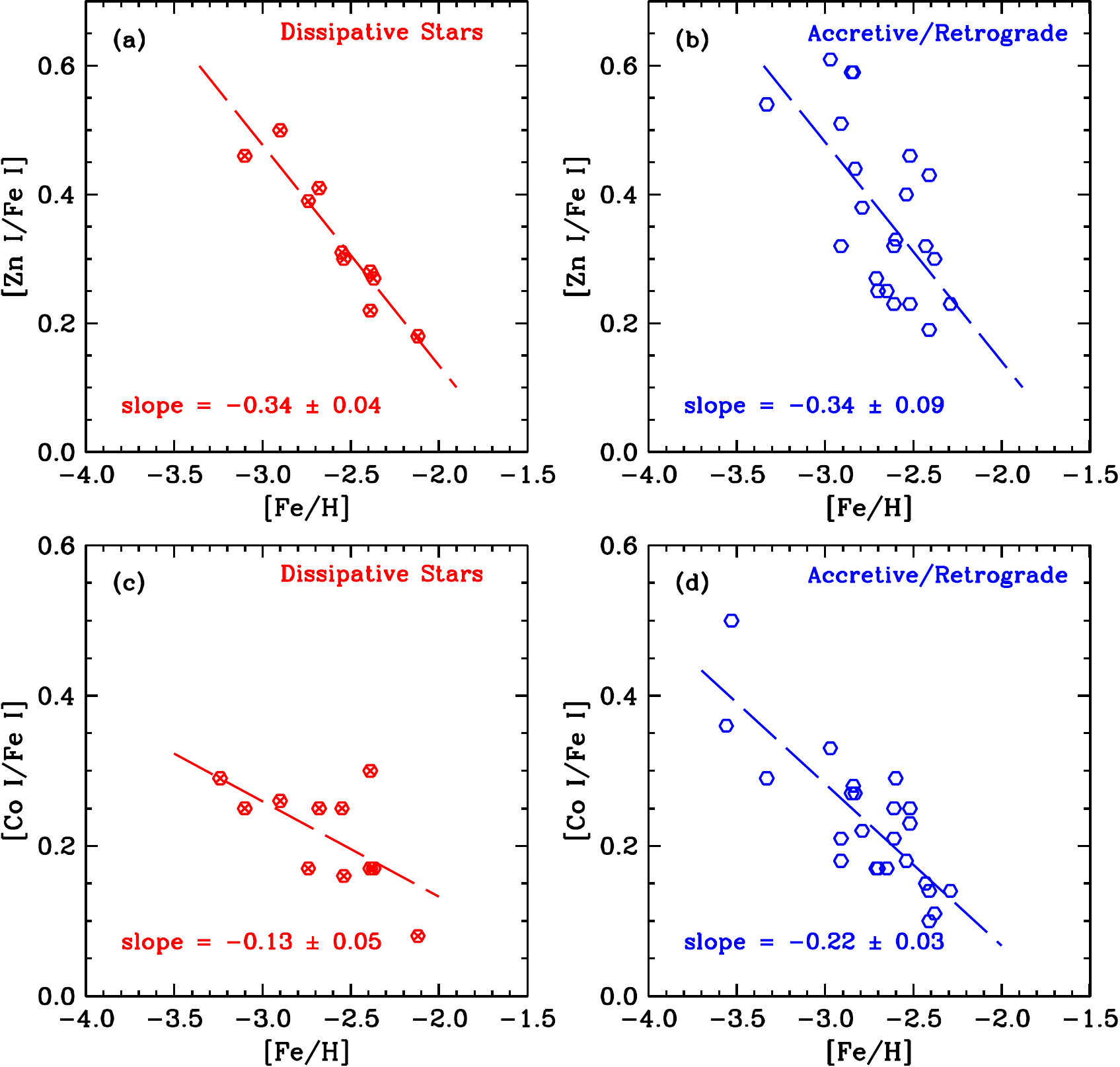}
\caption{
\label{fig12} \footnotesize
   Abundance ratios of Zn and Co with respect to [Fe/H] metallicity
   for 2 kinematic groups.
   Panel (a) shows the [\species{Zn}{i}/\species{Fe}{i}] values for the 
   dissipative stars and panel (b) shows them for the accretive retrograde 
   stars.
   Panels (c) and (d) repeat these plots for [\species{Co}{i}/\species{Fe}{i}].
   The dashed lines in each panel represent linear regression lines 
   through the data.  
}
\end{center}
\end{figure}
\vspace*{0.3in}

[Zn/Fe] is overabundant in all three metallicity groups displayed in 
Figure~\ref{fig6}.
This is also true to a lesser extent for [Co/Fe].  
We decided to examine the effects of stellar populations or Galactic 
components/origins on these trends.
In B11 their 117 stars were divided into groups based on their kinematic
properties.  
One stellar group was connected to the dissipative collapse of the Galaxy 
discussed by \cite{eggen62} includes stars from the classical thick disk 
and halo.  
Other stars are associated with a population that was accreted as proposed 
first by \cite{searle78} which are mainly halo stars.  
Some of those stars are on retrograde orbits.  
The details of the classifications can be found in B11.
Our sample of 37 stars here has 11 in the dissipative category and 25 
that are classified in the accretive population, of which 13 are also 
on retrograde orbits.
In Table~\ref{tab-zncokinem} we list the stars and their Zn and Co
abundance ratios in the two kinematic groups.
One program star not included in this table is BD~$-$13~3442; it lacked 
orbital/velocity information to be classified kinematically by B11.

In Figure~\ref{fig12} we show correlations of 
[\species{Zn}{i}/\species{Fe}{i}] with [Fe/H] for the dissipative component 
(panel a) and accretive component (panel b).
The mean trends with [Fe/H] are nearly identical.
But the star-to-star scatter is extremely small over an order of magnitude in 
metallicity for stars of the dissipative population.
Inspection of panel (b) reveals a significantly larger scatter in stars 
that were accreted into the Milky Way.
These representatives of nearby dwarf spheroidal galaxies have less uniform 
origins and come from different chemical environments.
This supposition is reinforced by the larger scatter in their 
[\species{Zn}{i}/\species{Fe}{i}] ratios.

The situation for Co is shown in Figure~\ref{fig12} panels (c) and (d).
A steeper slope with Fe for the accretive group is apparent, as well as
a somewhat larger spread in [\species{Co}{i}/\species{Fe}{i}].  
This could be attributed to the more diverse origin of this aggregate of stars. 
However, we repeat here our concerns on the apparent clash between neutral
and ionized Co species in our previous study \citep{cowan20}.
It is not clear how much to trust the [\species{Co}{i}/\species{Fe}{i}] 
results presented here.

Our understanding of the chemical and dynamical complexity of the Milky Way 
stellar halo has been revolutionized in recent years by the availability of 
large chemical abundance surveys and precise astrometric data (e.g., 
\citealt{naidu20}, \citealt{li21}, \citealt{limberg21}, \citealt{myeong22},
\citealt{horta23}).  
Our two-group categorization of stellar kinematics is appropriate here for simplicity and 
for our small sample
size, but a deeper analysis with larger data sets for Co and Zn will be
welcome. The results of this section suggest that there may be 
different abundance behavior among stars with different kinematics.  Future 
studies along these lines may provide additional insight into the nature of the 
supernova progenitors that enriched the stars in our sample.

\section{SUMMARY\label{sum}}

We have determined precise elemental abundance values for Mg, Ca, and 
Fe-group (Z = 21$-$30) elements for 37 warm, high gravity very metal-poor 
stars ($-$3.4~$\leq$~[Fe/H]~$\leq$~$-$2.1).
We have used high S/N and high spectral resolution echelle spectra which 
extend into the near-UV at 3050 \AA.  
We have measured 350+ spectral lines in each star of our sample.
Our analysis makes use of high quality lab data and includes hyperfine 
structure where needed.  
The abundances were found in LTE computations with various techniques for 
neutral and/or ionized states for 12 elements.  
For most elements there is little spread with small star-to-star scatter.  
The agreement in abundance from neutral and ionized species is very good for 
Ca, Ti, V, Cr, Mn, and Fe.

We have divided the stars into 3 metallicity groups and present the mean 
abundances for each set.  
For each group the trends in [X/Fe] are similar. 
The light Fe-peak elements, Sc, Ti, and V, are overabundant relative to Fe 
compared to solar ratios.  
Our study confirms and extends the apparent linkage between 
these overabundances previously found in other much smaller samples.
While Mn is underabundant, Co and especially Zn are overabundant.  
We have examined abundance correlations and find an indication of a new 
correlation between Zn and Ti.
This overabundance connection now includes Zn in the most metal-poor stars 
in our sample. 
There are also indications that both Ca and Zn may form in concert with Ti.

More observational determinations, as well as theoretical studies outside the 
scope of this paper will be needed to understand the production mechanism 
and history for these elements. 
Perhaps even more important will be NLTE abundance studies covering wide
ranges in model atmosphere parameters for large numbers of atomic transitions.
Our abundances derived from \species{Cr}{i}, \species{Mn}{i}, \species{Cu}{i},
and probably \species{Co}{i} point to clear limits to the reliability of
LTE analyses of Fe-group elements in metal-poor stars.

Employing the abundances data of our survey, we have examined 
the Galactic chemical evolution of a number of elements. 
While most [X/Fe] ratios either are constant with [Fe/H] metallicity
or change very modestly, increases in especially [Zn/Fe] are apparent in the 
lowest metallicity stars. 
These new results could provide important constraints on nucleosynthesis 
and SNe models early in Galactic history.

\begin{acknowledgments}
We thank Sanjana Curtis and Carla Fr{\" o}hlich for helpful comments and 
additional data for the comparison of nucleosynthesis models with our
abundances.
This work has been supported in part by: NASA grant NNX16AE96G (J.E.L.);
and National Science Foundation (NSF) grants AST-1516182, AST-1814512,
AST-2206050 (J.E.L. and E.D.H.), AST-1616040 (C.S.) and AST-1613536, 
AST-1815403/1815767/2205847 (IUR). 
Further assistance has come to IUR from NASA grant HST-AR-16630 from the 
Space Telescope Science Institute, which is operated by the Association of 
Universities for Research in Astronomy, Incorporated, under NASA contract 
NAS5-26555.
JJC and IUR were supported in part by the JINA Center for the Evolution of 
the Elements, supported by the NSF under Grant No. PHY-1430152.
This work has made use of data from the European Space Agency (ESA) mission 
Gaia (https://www.cosmos.esa.int/gaia), processed by the Gaia Data Processing 
and Analysis Consortium (DPAC, 
https://www.cosmos.esa.int/web/gaia/dpac/consortium). 
Funding for the DPAC has been provided by national institutions, in particular 
the institutions participating in the Gaia Multilateral Agreement.
Coauthor James E. Lawler passed away on January 29, 2023. 
The authors wish to acknowledge his invaluable contribution to this work 
and many others over a decades-long collaboration. His death represents a 
great loss not only for our collaboration, but for the entire laboratory 
astrophysics community. 
\end{acknowledgments}

\facility{Keck I (HIRES)}

\software{\textbf{linemake (https://github.com/vmplacco/linemake),
\citealt{placco21}, MOOG (\citealt{sneden73,sobeck11})}}


\clearpage
\bibliographystyle{apj}



\clearpage
\begin{center}
\begin{deluxetable}{crrrrrrrrr}
\tabletypesize{\footnotesize}
\tablewidth{0pt}
\tablecaption{Model Atmosphere Parameters\label{tab-models}}
\tablecolumns{10}
\tablehead{
\colhead{Star}                &
\colhead{\teff}               &
\colhead{\logg}               &
\colhead{\vmicro}             &
\colhead{[Fe/H]}              &
\colhead{$\sigma$}            &
\colhead{\#lines}             &
\colhead{[Fe/H]}              &
\colhead{$\sigma$}            &
\colhead{\#lines}             \\
\colhead{}                    &
\colhead{K}                   &
\colhead{}                    &
\colhead{\kmsec}              &
\colhead{\species{Fe}{i}}     &   
\colhead{\species{Fe}{i}}     &   
\colhead{\species{Fe}{i}}     &   
\colhead{\species{Fe}{ii}}    &   
\colhead{\species{Fe}{ii}}    &   
\colhead{\species{Fe}{ii}}    
}
\startdata
    G 64-12   &    6100  &    3.90  &    1.30  &     -3.60  &   0.11  &    113  &     -3.51  &   0.13  &   12 \\
    G 275-4   &    6050  &    4.20  &    1.10  &     -3.57  &   0.09  &     85  &     -3.49  &   0.11  &    7 \\
    G 64-37   &    6250  &    4.00  &    1.20  &     -3.34  &   0.09  &     98  &     -3.31  &   0.09  &   12 \\
  LP 831-70   &    6005  &    4.10  &    1.20  &     -3.28  &   0.09  &    121  &     -3.20  &   0.10  &   11 \\
   G 206-34   &    6000  &    4.00  &    1.20  &     -3.14  &   0.08  &    144  &     -3.05  &   0.11  &   15 \\
 BD +9 2190   &    6225  &    3.70  &    1.40  &     -3.01  &   0.10  &    118  &     -2.92  &   0.09  &   18 \\
BD $-$13 3442 &    6250  &    3.60  &    1.50  &     -2.94  &   0.09  &    115  &     -2.92  &   0.10  &   17 \\
BD +20 2030   &    6000  &    3.50  &    1.50  &     -2.95  &   0.10  &    138  &     -2.87  &   0.12  &   17 \\
  LP 815-43   &    6350  &    4.00  &    1.30  &     -2.97  &   0.10  &    117  &     -2.85  &   0.12  &   16 \\
  BD +3 740   &    6200  &    3.60  &    1.50  &     -2.95  &   0.11  &    137  &     -2.85  &   0.09  &   19 \\
 BD +1 2341   &    6350  &    4.00  &    1.40  &     -2.89  &   0.10  &    116  &     -2.81  &   0.10  &   17 \\
   LP 651-4   &    6275  &    4.00  &    1.30  &     -2.86  &   0.10  &    111  &     -2.82  &   0.08  &   12 \\
BD +26 2621   &    6275  &    4.10  &    1.40  &     -2.87  &   0.09  &    139  &     -2.79  &   0.09  &   19 \\
  LP 553-62   &    6200  &    3.70  &    1.40  &     -2.83  &   0.10  &    132  &     -2.74  &   0.11  &   17 \\
     G 92-6   &    6275  &    3.90  &    1.40  &     -2.76  &   0.10  &    144  &     -2.71  &   0.10  &   18 \\
    G 26-12   &    5950  &    3.75  &    1.25  &     -2.73  &   0.09  &    144  &     -2.69  &   0.10  &   17 \\
  LP 635-14   &    6150  &    3.55  &    1.40  &     -2.72  &   0.10  &    148  &     -2.67  &   0.11  &   21 \\
     G 4-37   &    6200  &    4.00  &    1.30  &     -2.71  &   0.11  &    153  &     -2.64  &   0.10  &   17 \\
   G 181-28   &    5950  &    4.00  &    1.25  &     -2.68  &   0.10  &    163  &     -2.62  &   0.11  &   16 \\
    G 88-10   &    6100  &    4.00  &    1.25  &     -2.64  &   0.09  &    155  &     -2.58  &   0.12  &   19 \\
   G 108-58   &    5800  &    4.30  &    1.20  &     -2.63  &   0.11  &    146  &     -2.59  &   0.09  &   16 \\
BD +24 1676   &    6300  &    3.70  &    1.50  &     -2.62  &   0.10  &    162  &     -2.58  &   0.09  &   21 \\
BD $-$4 3208  &    6200  &    3.65  &    1.45  &     -2.59  &   0.10  &    154  &     -2.51  &   0.09  &   20 \\
BD +13 3683   &    5500  &    3.10  &    1.25  &     -2.54  &   0.11  &    147  &     -2.54  &   0.08  &   16 \\
   G 126-52   &    6200  &    3.80  &    1.30  &     -2.58  &   0.10  &    159  &     -2.49  &   0.09  &   20 \\
    G 201-5   &    6150  &    3.90  &    1.20  &     -2.53  &   0.10  &    162  &     -2.51  &   0.10  &   21 \\
    G 59-24   &    6100  &    4.30  &    1.30  &     -2.55  &   0.10  &    146  &     -2.49  &   0.12  &   15 \\
 BD +2 3375   &    6025  &    3.90  &    1.25  &     -2.43  &   0.10  &    157  &     -2.43  &   0.09  &   17 \\
   LTT 1566   &    6125  &    3.90  &    1.20  &     -2.42  &   0.11  &    169  &     -2.40  &   0.08  &   19 \\
    G 75-56   &    6100  &    3.70  &    1.25  &     -2.44  &   0.09  &    160  &     -2.38  &   0.08  &   18 \\
BD $-$10 388  &    6350  &    4.00  &    1.50  &     -2.43  &   0.11  &    171  &     -2.35  &   0.08  &   20 \\
BD $-$14 5850 &    5775  &    4.10  &    1.25  &     -2.35  &   0.12  &    288  &     -2.42  &   0.16  &   25 \\
  LP 752-17   &    5950  &    3.20  &    1.50  &     -2.39  &   0.09  &    150  &     -2.37  &   0.09  &   19 \\
BD +36 2964   &    6150  &    3.70  &    1.35  &     -2.39  &   0.09  &    156  &     -2.34  &   0.08  &   19 \\
   G 130-65   &    6100  &    3.90  &    1.35  &     -2.29  &   0.10  &    154  &     -2.28  &   0.09  &   19 \\
    G 20-24   &    6200  &    3.60  &    1.30  &     -2.14  &   0.10  &    188  &     -2.10  &   0.10  &   25 \\
BD +51 1696   &    5695  &    4.50  &    1.00  &     -1.40  &   0.12  &    134  &     -1.44  &   0.08  &   17 \\
\enddata                                                      
\end{deluxetable}                                             
\end{center} 

\begin{deluxetable}{lrcrrr}
\tablewidth{0pt}
\tablecaption{Abundances for Individual Stars\label{tab-abunds}}
\tablecolumns{6}
\tablehead{
\colhead{Star}                      & 
\colhead{[Fe/H]\tablenotemark{a}}   &
\colhead{Species\tablenotemark{b} } &
\colhead{[X/Fe]}                    & 
\colhead{$\sigma$}                  &
\colhead{\#lines}          
}
\startdata
G64-12 & $-$3.56 & \species{Mg}{i}        &    0.57 &         0.10 &  7 \\
G64-12 & $-$3.56 & \species{Ca}{i}        &    0.60 &         0.05 & 13 \\
G64-12 & $-$3.56 & \species{Ca}{ii}       &    0.54 &         0.04 &  5 \\
G64-12 & $-$3.56 & \species{Sc}{ii}       &    0.11 &         0.04 & 11 \\
G64-12 & $-$3.56 & \species{Ti}{i}        &    0.65 &         0.07 &  6 \\
G64-12 & $-$3.56 & \species{Ti}{ii}       &    0.40 &         0.12 & 62 \\
G64-12 & $-$3.56 & \species{V}{i}         & \nodata & \nodata & \nodata \\
G64-12 & $-$3.56 & \species{V}{ii}        &    0.17 &         0.07 &  6 \\
G64-12 & $-$3.56 & \species{Cr}{i}        &    0.17 &         0.07 &  6 \\
G64-12 & $-$3.56 & \species{Cr}{i} no 0eV & $-$0.13 &         0.08 &  6 \\
G64-12 & $-$3.56 & \species{Cr}{ii}       & $-$0.06 &         0.09 &  2 \\
\enddata
\tablenotetext{a}{[Fe/H] is the mean of [\species{Fe}{i}/H] and
                  [\species{Fe}{ii}/H] values.}
\tablenotetext{b}{The designation ``no 0eV'' means the abundance computed 
                  for this speices without inclusion of lines arising from the 
                  ground state.}
\tablecomments{This table is available in its entirety in machine-readable 
form.}
\end{deluxetable}

\begin{deluxetable}{crrrrrr}
\tablewidth{0pt}
\tablecaption{Mean Species Abundance Ratios\label{tab-specmean}}
\tablecolumns{7}
\tablehead{
\colhead{Element}                    &
\colhead{$\langle$[X/Fe]$\rangle$}   &
\colhead{$\sigma$}                   &
\colhead{\#stars}                    &
\colhead{$\langle$[X/Fe]$\rangle$}   &
\colhead{$\sigma$}                   &
\colhead{\#stars}                    \\
\colhead{}                           &
\colhead{I}                          &
\colhead{I}                          &
\colhead{I}                          &
\colhead{II}                         &
\colhead{II}                         &
\colhead{II}                         
}
\startdata
Mg &    0.44 &    0.09 &      37 & \nodata & \nodata & \nodata \\
Ca &    0.44 &    0.09 &      37 &    0.42 &    0.11 &      25 \\
Sc & \nodata & \nodata & \nodata &    0.10 &    0.09 &      37 \\
Ti &    0.46 &    0.10 &      37 &    0.35 &    0.06 &      37 \\
V  &    0.13 &    0.09 &       7 &    0.17 &    0.08 &      37 \\
Cr\tablenotemark{a}
   & $-$0.03 &    0.06 &      37 &    0.03 &    0.04 &      37 \\
Mn\tablenotemark{a}
   & $-$0.28 &    0.06 &      19 & $-$0.27 &    0.09 &      37 \\
Co &    0.21 &    0.02 &      37 & \nodata & \nodata & \nodata \\
Ni &    0.02 &    0.07 &      37 & \nodata & \nodata & \nodata \\
Cu & $-$0.71 &    0.08 &      36 & \nodata & \nodata & \nodata \\
Zn &    0.35 &    0.13 &      34 & \nodata & \nodata & \nodata \\
\enddata
\tablenotetext{a}{these species means were computed without inclusion
                  of the 0~eV resonance lines}

\end{deluxetable}

\begin{deluxetable}{crrrr}
\tablewidth{0pt}
\tablecaption{Mean Element Abundance Ratios in Three Metallicity Bins\label{tab-elmean}}
\tablecolumns{5}
\tablehead{
\colhead{Element}                    &
\colhead{$\langle$[X/Fe]$\rangle$}   &
\colhead{$\langle$[X/Fe]$\rangle$}   &
\colhead{$\langle$[X/Fe]$\rangle$}   &
\colhead{$\langle$[X/Fe]$\rangle$}   \\
\colhead{$\langle$[Fe/H]$\rangle$}   &
\colhead{all}                        &
\colhead{$-$3.09}                    &
\colhead{$-$2.67}                    &
\colhead{$-$2.32}                    
}
\startdata
Mg &    0.44 &    0.44 &    0.45 &    0.43 \\
Ca &    0.43 &    0.44 &    0.44 &    0.42 \\
Sc &    0.10 &    0.14 &    0.07 &    0.08 \\
Ti &    0.40 &    0.45 &    0.39 &    0.37 \\
V  &    0.17 &    0.22 &    0.12 &    0.16 \\
Cr &    0.00 &    0.02 & $-$0.01 &    0.00 \\
Mn & $-$0.27 & $-$0.28 & $-$0.29 & $-$0.26 \\
Fe &    0.00 &    0.00 &    0.00 &    0.00 \\
Co &    0.21 &    0.29 &    0.21 &    0.15 \\
Ni &    0.02 &    0.02 &    0.01 &    0.03 \\
Cu & $-$0.71 & $-$0.69 & $-$0.72 & $-$0.72 \\
Zn &    0.35 &    0.50 &    0.32 &    0.28 \\
\enddata

\end{deluxetable}

\begin{deluxetable}{crrr}
\tablewidth{0pt}
\tablecaption{Co and Zn in Kinematic Groups\label{tab-zncokinem}}
\tablecolumns{4}
\tablehead{
\colhead{Star\tablenotemark{a}}             &
\colhead{[Fe/H]}                            &
\colhead{[\species{Co}{i}/\species{Fe}{i}]} &
\colhead{[\species{Zn}{i}/\species{Fe}{i}]}   
}
\startdata
\multicolumn{4}{c}{Dissipative Group}    \\
LP 831-70      & $-$3.24 &   0.29 & \nodata \\
G 206-34       & $-$3.10 &   0.25 &  0.46   \\
BD+3-740       & $-$2.90 &   0.26 &  0.50   \\
G 92-6         & $-$2.74 &   0.17 &  0.39   \\
G 4-37         & $-$2.68 &   0.25 &  0.41   \\
BD $-$4-3208   & $-$2.55 &   0.25 &  0.31   \\
BD +13 3683    & $-$2.54 &   0.16 &  0.30   \\
BD $-$10-388   & $-$2.39 &   0.30 &  0.28   \\
BD $-$14-5850  & $-$2.39 &   0.17 &  0.22   \\
BD +36-2964    & $-$2.37 &   0.17 &  0.27   \\
G 20-24        & $-$2.12 &   0.08 &  0.18   \\           
\multicolumn{4}{c}{Accretive Group}      \\
G 64-12        & $-$3.56 &   0.36 & \nodata \\
G 275-4        & $-$3.53 &   0.50 & \nodata \\
G 64-37        & $-$3.33 &   0.29 &    0.54 \\
BD +9-2190     & $-$2.97 &   0.33 &    0.61 \\
BD +20-2030    & $-$2.91 &   0.18 &    0.51 \\
LP 815-43      & $-$2.91 &   0.21 &    0.32 \\
BD +11-2341    & $-$2.85 &   0.27 &    0.59 \\
LP 651-4       & $-$2.84 &   0.28 &    0.59 \\
BD +26-2621    & $-$2.83 &   0.27 &    0.44 \\
LP 553-62      & $-$2.79 &   0.22 &    0.38 \\
G 26-12        & $-$2.71 &   0.17 &    0.27 \\
LP 635-14      & $-$2.70 &   0.17 &    0.25 \\
G 181-21       & $-$2.65 &   0.17 &    0.25 \\
G 88-10        & $-$2.61 &   0.25 &    0.32 \\
G 108-58       & $-$2.61 &   0.21 &    0.23 \\
BD +24-1676    & $-$2.60 &   0.29 &    0.33 \\
G 126-52       & $-$2.54 &   0.18 &    0.40 \\
G 201-5        & $-$2.52 &   0.23 &    0.23 \\
G 59-24        & $-$2.52 &   0.25 &    0.46 \\
BD +2-3375     & $-$2.43 &   0.15 &    0.32 \\
LTT 1566       & $-$2.41 &   0.10 &    0.19 \\
G 75-56        & $-$2.41 &   0.14 &    0.43 \\
LP 752-17      & $-$2.38 &   0.11 &    0.30 \\
G 130-65       & $-$2.29 &   0.14 &    0.23 \\
\enddata

\tablenotetext{a}{excluding the higher metallicity star BD +51 1696,
                  and BD~$-$13~3442 for which kinematic group membership
                  could not be determined by B11}

\end{deluxetable}

\end{document}